\documentclass[twocolumn]{aastex63}

\usepackage{amsmath}
\usepackage{colortbl}

\newcommand{\result}[1]{\textcolor{black}{#1}}

\begin{document}

\title{Apples and Oranges: Comparing black holes in X-ray binaries and gravitational-wave sources}
\author{Maya Fishbach}
\altaffiliation{NASA Hubble Fellowship Program Einstein Postdoctoral Fellow}
\affiliation{Center for Interdisciplinary Exploration and Research in Astrophysics (CIERA) and Department of Physics and Astronomy,
Northwestern University, 1800 Sherman Ave, Evanston, IL 60201, USA}
\author{Vicky Kalogera}
\affiliation{Center for Interdisciplinary Exploration and Research in Astrophysics (CIERA) and Department of Physics and Astronomy,
Northwestern University, 1800 Sherman Ave, Evanston, IL 60201, USA}

\begin{abstract}
The component black holes (BHs) observed in gravitational-wave (GW) binary black hole (BBH) events tend to be more massive and slower spinning than those observed in black hole X-ray binaries (BH-XRBs).
Without modeling their evolutionary histories, we investigate whether these apparent tensions in the BH populations can be explained by GW observational selection effects alone.
We find that this is indeed the case for the discrepancy between BH masses in BBHs and the observed high-mass X-ray binaries (HMXBs), when we account for statistical uncertainty from the small sample size of just three HMXBs. On the other hand, the BHs in observed low-mass X-ray binaries (LMXBs) are significantly lighter than the astrophysical BBH population, but this may just be due to a correlation between component masses in a binary system. Given their light stellar companions, we expect light BHs in LMXBs. The observed spins in HMXBs and LMXBs, however, are in tension with the inferred BBH spin distribution at the $>99.9\%$ level.
We discuss possible scenarios behind the significantly larger spins in observed BH-XRBs. One possibility is that a small subpopulation (conservatively $<30\%$) of BBHs have rapidly spinning primary components, indicating that they may have followed a similar evolutionary pathway to the observed HMXBs. In LMXBs, it has been suggested that BHs can spin up by accretion. If LMXB natal spins follow the BBH spin distribution, we find LMXBs must gain an average dimensionless spin of $0.47^{+0.10}_{-0.11}$, but if their natal spins follow the observed HMXB spins, the average spinup must be $<0.03$.

\end{abstract}

\section{Introduction}
\label{sec:intro}
While very massive stars are rare and short-lived, we can piece together their histories by observing their remnants: stellar-mass black holes (BHs).
Before the first detection of gravitational waves (GWs;~\citealt{2016PhRvL.116f1102A}), stellar-mass BHs were discovered in X-ray binary systems~\citep{2006csxs.book..623T,2006ARA&A..44...49R,2014SSRv..183..223C,2021NewAR..9301618M}. Black hole X-ray binaries (BH-XRBs) consist of a BH accreting from a non-degenerate binary companion, thereby powering X-ray emission. There are two classifications of BH-XRBs, characterized by the mass of the donor star and the type of accretion: low-mass X-ray binaries (LMXBs), powered by Roche lobe accretion from a typically low-mass ($\lesssim 2$--$3\,M_\odot$) star, and high-mass X-ray binaries (HMXBs), powered by wind-fed accretion from a typically high-mass ($\gtrsim 5\,M_\odot$) star. There are currently $\sim 30$ known BHs in BH-XRB systems.
Meanwhile, over the last few years, the discovery rate of BHs has accelerated dramatically with the advent of GW astronomy. Advanced LIGO~\citep{2015CQGra..32g4001L} and Virgo~\citep{2015CQGra..32b4001A} have confidently discovered $\sim 50$ binary black holes (BBHs), or equivalently $\sim 100$ component BHs, in their first two catalogs, up to the most recent catalog GWTC-2.1~\citep{2016PhRvX...6d1015A,2019PhRvX...9c1040A,2021PhRvX..11b1053A,2021arXiv210801045T}, with $\sim 5$ additional confident BBHs found in public data~\citep{2019PhRvD.100b3007Z,2020PhRvD.101h3030V,2019ApJ...872..195N,2020ApJ...891..123N,2021arXiv210509151N,2020PhRvD.102l3022R}. In addition to BH-XRBs and BBHs, there is a small but growing number of noninteracting BHs discovered by gravitational microlensing~\citep{2016ApJ...830...41L,2020A&A...636A..20W} and radial velocity measurements~\citep{2018MNRAS.475L..15G,2019Sci...366..637T}. In this work, we focus on comparing the properties of BHs in BH-XRB and BBH systems. We use the term ``BBH" to refer to the population of double BH systems that merge within a Hubble time. 

The comparison between LMXB, HMXB and BBH populations is complicated by the presence of \emph{astrophysical} and \emph{observational} selection effects. Although the BHs in LMXBs, HMXBs and BBHs all likely originate from massive stars, they undergo different evolutionary pathways (giving rise to astrophysical selection effects) and are subject to different detection criteria (giving rise to observational selection effects). 

As an example of a possible astrophysical selection effects, LMXBs will never evolve into double BH systems, merging or non-merging, because their donor stars are too light. Meanwhile, only a subset of HMXBs go on to become merging BBH systems; this subset does not include any of the observed HMXBs~\citep{2011ApJ...742L...2B,2012arXiv1208.2422B,2021ApJ...908..118N}. In the other direction, some BBH systems may be assembled through dynamical interactions in dense stellar environments, so that perhaps not all BBH systems evolve from HMXBs. If the probability that a BH-XRB evolves into a merging BBH, or inversely, the probability that a BBH has a BH-XRB progenitor~\citep{2019ApJ...878L...1P,2021A&A...645A...5S,Liotine}, correlates with BH mass or spin, the astrophysical mass and spin distributions will differ between the populations. For example, in order to evolve into a BBH system, both components of a BH-XRB system must be sufficiently massive. Thus, if the two component masses in a binary are positively correlated, we expect relatively light BHs in systems including LMXBs with a relatively light companion. We return to this example in Section~\ref{sec:LMXBmass}. Predicting the presence of astrophysical selection effects and assessing their significance requires an astrophysical model for the formation and evolution of all relevant populations; for such an astrophysical study see~\citet{Liotine}. 

Observational selection effects arise when the masses and spins of the BH(s) in the system affect its detection probability. For BBHs, the GW signal as a function of the component masses, spins, and merger redshift is known, and so the detection probability is straightforward to quantify by running the detection pipelines over simulated signals injected into GW detector data. In general, heavier BBHs with large aligned spins emit a louder GW signal, so the GW observational selection effects favor large masses and, to a lesser degree, large aligned spins.
In contrast, while the observed emission of BH-XRBs depends on the BH mass and spin, it also depends on other characteristics that require astrophysical modeling and cannot be cleanly extracted from the data~\citep{Liotine,Siegel}. 
For example, several proposed explanations for the small BH masses in BH-XRBs invoke observational selection effects, including the explanation that observed HMXBs preferentially formed at high metallicities, since BH-XRBs are detectable only in the local universe, or the explanation that observed BH-XRBs preferentially received large kicks~\citep{2021arXiv210403596J}. These explanations rely on theory, namely the expected anti-correlation between the progenitor metallicity and the BH mass~\citep{2021MNRAS.504..146V}, or the prediction that low-mass BHs receive larger kicks, respectively.

The goal of this paper is to compare the masses and spins of BHs in BH-XRBs and BBHs without model systematics, and so we are limited to comparing the \emph{observed} population of BH-XRBs and the \emph{astrophysical} population of BBHs.
Therefore, while we account for GW observational selection effects to accurately infer the astrophysical BBH mass and spin distributions~\citep{2019MNRAS.486.1086M,2021ApJ...913L...7A}, we do not model these selection effects in BH-XRB systems, as the latter relies on uncertain theoretical calculations. The statistical tensions we identify between the astrophysical BBH and the observed BH-XRB populations may stem from an astrophysical difference in evolutionary histories, an observational detection bias affecting BH-XRBs, or a combination of the two, and we discuss several possible explanations.

The remainder of the paper is structured as follows. We introduce the BH-XRB and GW datasets and the statistical population inference in Section~\ref{sec:data}. In Section~\ref{sec:masses}, we evaluate the consistency between the measured BH masses in LMXBs and HMXBs and the inferred BBH mass distribution. 
We turn to the BH spin distribution in Section~\ref{sec:spins}, where we quantify the clear tension between {the measured} BH spins in BBHs and BH-XRBs. Motivated by the rapid BH spins {reported} for HMXBs, we fit new population models to the BBH spin distribution to investigate the possible presence of a sub-population with large primary spins.

\section{Data and methods}
\label{sec:data}

\begin{table}[]
    \centering
    \begin{tabular}{l c c}
    \hline
    {\bf Name} & {\bf BH mass} $[M_{\odot}]$ & {\bf BH spin}  \\
    \hline \hline
\rowcolor{gray!20}[5pt][5pt]
M33 X-7 & $15.65^{+1.45}_{-1.45}$ & $0.84^{+0.05}_{-0.05}$ \\
\rowcolor{gray!20}[5pt][5pt]
Cygnus X-1 & $21.20^{+2.20}_{-2.20}$ & $>0.983$ \\
\rowcolor{gray!20}[5pt][5pt]
LMC X-1 & $10.90^{+1.40}_{-1.40}$ & $0.92^{+0.05}_{-0.07}$\footnote{Reflection spectroscopy spin: $> 0.55$} \\
\hline
Swift J1357.2–0933 & $> 8.9$ & --  \\
XTE J1650–500 & $<7.3$ & $0.79^{+0.01}_{-0.01}$  \\
XTE J1118+480 & $7.55^{+0.65}_{-0.65}$ & --  \\
XTE J1859+226 (V406 Vul) & $> 5.42$ & --  \\
SAX J1819.3-2525 (V4641 Sgr) & $6.40^{+0.60}_{-0.60}$ & --  \\
XTE J1550-564 & $11.70^{+3.89}_{-3.89}$ & $0.34^{+0.37}_{-0.45}$\footnote{Reflection spectroscopy spin: 0.33--0.77} \\
GRO J1655–40 (N. Sco 94) & $6.00^{+0.40}_{-0.40}$ & $0.70^{+0.10}_{-0.10}$\footnote{Reflection spectroscopy spin: $> 0.9$} \\
GRS 1009-45 (N. Vel 93) & $> 4.4$ & --  \\
GRS 1915+105 & $12.00^{+2.00}_{-2.00}$ & $0.88^{+0.06}_{-0.13}$\footnote{Continuum fitting spin: $>0.95$}  \\
GRO J0422+32 & $8.50^{+6.50}_{-6.50}$ & --  \\
GRS 1124-684 (N. Mus 91) & $5.65^{+1.85}_{-1.85}$ & $0.63^{+0.16}_{-0.19}$ \\
GS 2023+338 (V404 Cyg) & $9.00^{+0.20}_{-0.60}$ & $>0.92$ \\
GS 2000+251 (QZ Vul) & $7.15^{+1.65}_{-1.65}$ & --  \\
GS 1354-64 (BW Cir) & $>6.9$ & $>0.98$ \\
H 1705-250 (N. Oph 77) & $6.40^{+1.50}_{-1.50}$ & --  \\
3A0620–003 & $6.60^{+0.30}_{-0.30}$ & $0.12^{+0.19}_{-0.19}$   \\
1H J1659-487 (GX 339-4) & $>6.0$ & $>0.95$ \\
4U 1543–475 (IL Lup) & $9.40^{+1.00}_{-1.00}$ & $0.67^{+0.15}_{-0.08}$\footnote{Continuum fitting spin: $0.8 \pm 0.1$} \\
GRS 1716-249 & $6.45^{+1.55}_{-1.55}$ & $>0.92$ \\
LMC X-3 & $6.98^{+0.56}_{-0.56}$ & $0.25^{+0.20}_{-0.29}$ \\
XTE J1652–453 & --  & $0.45^{+0.02}_{-0.02}$ \\
XTE J1752–223 & --  & $0.92^{+0.06}_{-0.06}$ \\
Swift J1910.2–0546 & --  & $>0.32$ \\
MAXI J1836–194 & --  & $0.88^{+0.03}_{-0.03}$ \\
XTE J1908+094 & --  & $0.75^{+0.09}_{-0.09}$ \\
Swift J1753.5–0127 & --  & $0.76^{+0.11}_{-0.15}$ \\
4U 1630–472 & --  & $>0.97$ \\
SAX J1711.6–3808 & --  & $0.60^{+0.20}_{-0.40}$ \\
EXO 1846–031 & --  & $>0.99$ \\ \hline

    \end{tabular}
    \caption{BH-XRB systems used in this work with their name (first column), BH mass (second column) and dimensionless BH spin magnitude (third column). We use the measurements collected by~\citet{2015yCat..35870061C,2016A&A...587A..61C,2019ApJ...870L..18Q,doi:10.1146/annurev-astro-112420-035022}. We group the systems by HMXB (gray-filled; above the horizontal line) and LMXB (below the horizontal line). For systems where both reflection spectroscopy and continuum fitting measurements of the BH spin are available, we include one value in the main table and the other value in the footnotes (see main text).}
    \label{tab:BH-XRB}
\end{table}

For our BBH sample, we use the 44 GWTC-2 events detected with a false-alarm-rate FAR $< 1/$ year and exclude events with secondary masses $m_2 < 3\,M_\odot$~\citep{2021PhRvX..11b1053A}. In Sections~\ref{sec:masses}--\ref{sec:spinup}, we adopt the fits to the BBH mass and spin distribution presented in~\citet{2021ApJ...913L...7A}, using the ``Default" spin model~\citep{2019PhRvD.100d3012W,2017PhRvD..96b3012T} and the ``Power Law + Peak" mass model~\citep{2018ApJ...856..173T}.\footnote{These population fits are available at \url{https://dcc.ligo.org/LIGO-P2000434/public}.} In Section~\ref{sec:spinsa1large}, we fit a new model to the BBH population, carrying out a hierarchical Bayesian inference~\citep{2004AIPC..735..195L,2019MNRAS.486.1086M,2019PASA...36...10T,2020arXiv200705579V} using the publicly available posterior samples and sensitivity estimates~\citep{ligo_scientific_collaboration_and_virgo_2021_748570}. 

Table~\ref{tab:BH-XRB} shows measured BH masses (in solar masses) and BH spin magnitudes (dimensionless spin parameter) of a sample of observed HMXBs and LMXBs collected by several references~\citep[][see observational references therein]{2015yCat..35870061C,2016A&A...587A..61C,2019ApJ...870L..18Q,doi:10.1146/annurev-astro-112420-035022}, including the recently updated mass of Cygnus X-1~\citep{2021Sci...371.1046M}. The horizontal line separates the HMXBs (gray-filled; above the line) from the LMXBs (below the line). We exclude the two HMXB systems with Wolf-Rayet companions, NGC 300 X-1 and IC 10 X-1, from our analysis as their measurements are less reliable~\citep{2015MNRAS.452L..31L}. In cases where only a range of mass values is reported in~\citet{2016A&A...587A..61C}, we approximate the measurement by the midpoint and equal-sized plus/minus error bars. The BH-XRB spins collected here are measured by two main techniques: disk continuum fitting and reflection spectroscopy~\citep{2015PhR...548....1M,doi:10.1146/annurev-astro-112420-035022}. For systems where only one technique is available (the continuum fitting method requires an independent measurement of the BH mass, for example), we report the available measurement from~\citet{doi:10.1146/annurev-astro-112420-035022}. For systems where both techniques provide a spin measurement, we take the \emph{smaller} of the two reported spins, so that our estimate of the tension between the BH-XRB and BBH spin distributions are always as conservative as possible. The exception to this is if two measurements are available, but one is only an upper or lower limit. In this case, we choose the measurement that reports a central value and credible interval. The alternative spin measurement is reported in the footnotes of Table~\ref{tab:BH-XRB}. {For systems where both the reflection spectroscopy and continuum fitting methods are available, their spin measurements are generally in good agreement~\citep{doi:10.1146/annurev-astro-112420-035022}.} We report spin \emph{magnitudes} in the range $0 < a <1$, rather than the BH spin component projected along the orbital angular momentum axis. In particular, the reported spin of Swift J1910.2-0546 is retrograde, and reported as $a<-0.32$~\citep{2013ApJ...778..155R}, but here we consider it to be $a > 0.32$. We do not account for possible systematic uncertainties affecting BH-XRB spin measurements~\citep[e.g.][]{2018ApJ...855..120T,2021MNRAS.500.3640S,2021MNRAS.504.3424F}. {We expect these systematic uncertainties to contribute an additional $\lesssim 0.1$ to the error budget of individual BH-XRB measurements, which does not qualitatively affect our results.} 

When we fit the BH-XRB spin distributions in Section~\ref{sec:spins}, in the absence of posterior samples for each system, we approximate the likelihood for each system as Gaussian, centered on the reported median spin parameter, with the standard deviation chosen so that the central 90\% probability matches the width of the reported error range. We truncate the Gaussian on the physical range $0 < a < 1$. In the case where only an upper or lower limit is reported, we approximate the likelihood as a broad half-Gaussian centered on the limit. We choose the standard deviation of the underlying Gaussian, centered at the reported lower (upper) limit, so that it encloses the full physical range between the lower (upper) limit and 1 (0) at 90\% probability. We then truncate the Gaussian between lower (upper) limit and 1 (0). We then use our approximate likelihoods for each system in a standard hierarchical Bayesian analysis framework to measure the parameters that govern the observed spin distribution. Unlike the BBH case, we do not account for BH-XRB observational selection effects. 

\section{Black hole mass}
\label{sec:masses}
We start by comparing the mass distribution between BHs in BBH and HMXB systems (Section~\ref{sec:HMXBmass}) and then between BBH and LMXB systems (Section~\ref{sec:LMXBmass}).
When comparing BH masses between BBH and BH-XRB systems, we must consider the subtlety that GW observations measure the \emph{two-dimensional} mass distribution $p(m_1, m_2)$ for both component BHs in BBH systems, where the primary mass $m_1$ (secondary mass $m_2$) is defined as the more (less) massive component. Meanwhile, BH-XRB systems only contain one BH, which is expected to have evolved from the initially more massive star in the binary, and will usually, although not always, become the primary BH if the system later evolves into a BBH. In the following, we compare the marginal primary mass distribution $p(m_1)$ for BBHs, as inferred by~\citet{2021ApJ...913L...7A}, to the masses of BHs in BH-XRBs. By considering BBH primary masses, we exaggerate any tension between the GW and X-ray distributions, because the BHs in BBH systems are apparently more massive than those in BH-XRBs. Even so, in the following we show that, under certain assumptions, the tensions between the BBH and BH-XRB distributions are small.

\subsection{Black holes in high-mass X-ray binaries have similar masses to binary black holes}
\label{sec:HMXBmass}
\begin{figure}
    \centering
    \includegraphics[width = 0.5\textwidth]{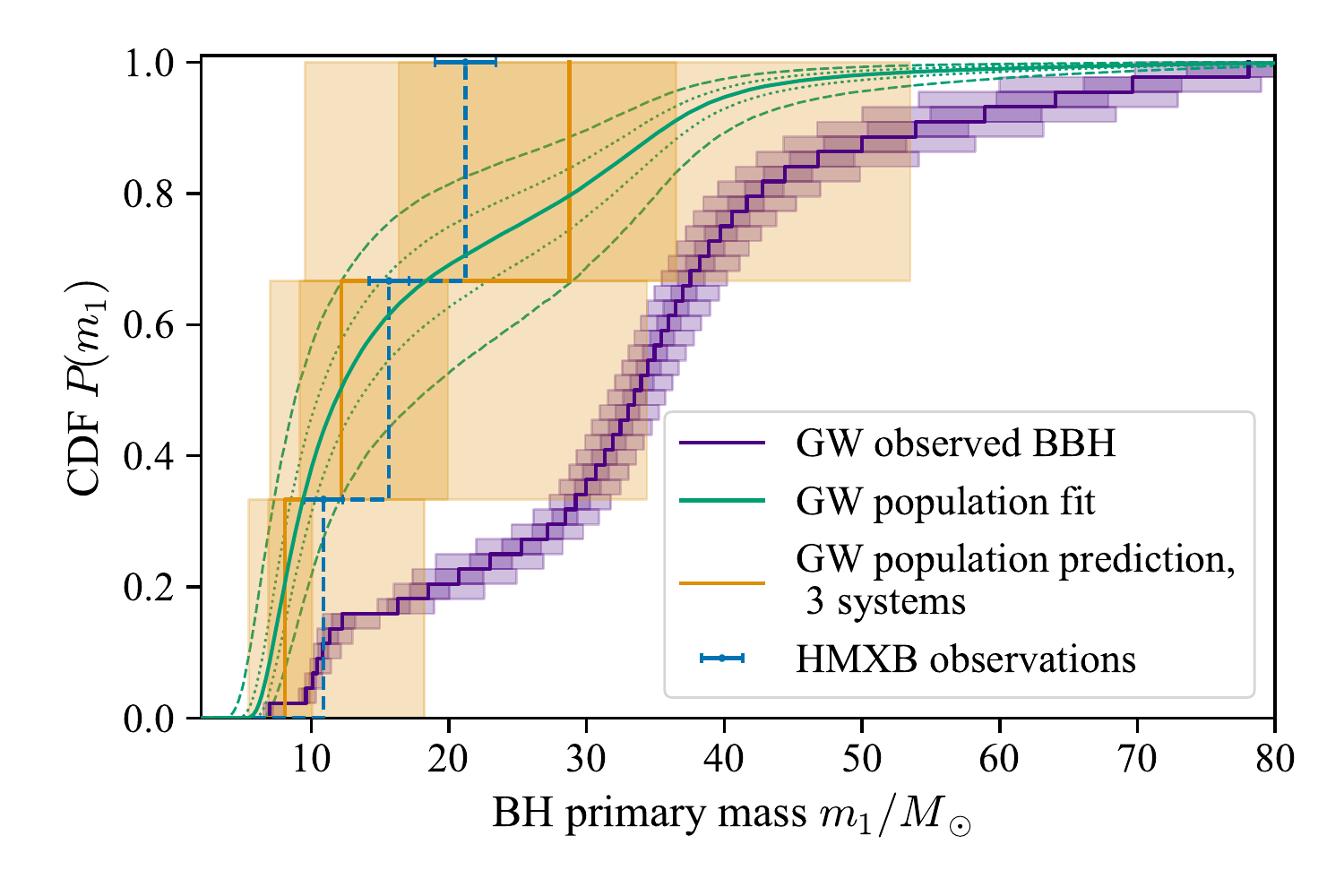}
    \caption{Empirical CDF for the three observed HMXB BH masses (blue error bars), compared to the GW population inference. Purple bands show the CDF of the 44 \emph{observed} BBH primary masses, before correcting for GW selection effects. Correcting for GW selection effects, the fit to the \emph{astrophysical} BBH primary mass distribution is shown by the green, unfilled bands. Finally, we add in small-sample-size Poisson uncertainty, showing the predicted CDF of \emph{three random draws} from the astrophysical BBH primary mass distribution (orange bands). Here and throughout all figures, shaded regions show 50\% and 90\% credible intervals. We see that while the observed BBH distribution (purple) skews to much larger masses, accounting for GW selection effects (green) and Poisson uncertainty (orange) brings the two distributions into agreement within statistical uncertainties, so that there is no evidence that BHs in the observed HMXBs follow a different mass distribution from the astrophysical BBH population.}
    \label{fig:mass_comparison_HMXB}
\end{figure}

Figure~\ref{fig:mass_comparison_HMXB} compares the BH masses of the three confident HMXB systems from Table~\ref{tab:BH-XRB} (blue) to the \emph{observed} and \emph{astrophysical} BBH primary mass distribution from GWTC-2 (purple, green and orange). {The observed distribution describes detected BBH events, subject to GW selection effects, while the astrophysical distribution describes all merging BBH sources in the Universe, regardless of their detectability.}
In blue, we show the empirical cumulative distribution function (CDF) of the HMXB BH mass measurements, ordered from lightest to heaviest BH by the median measured mass, with measurement uncertainty denoted by the error bars. 
In purple, we show the primary masses of the 44 confident BBH events from GWTC-2, before correcting for GW selection effects. We see that the observed BBH primary masses are larger than those observed in HMXBs. Correcting for GW selection effects that favor massive BBHs, the fit to the astrophysical primary mass distribution from~\citet{2021ApJ...913L...7A} is shown by the green, unfilled bands. 
In orange, we account for Poisson uncertainty, and show the predicted CDF of three random draws from the astrophysical mass distribution shown in green.
To construct the orange bands, we draw a set of three random masses from each primary mass distribution within the green unfilled bands. Each set of three random mass draws, {ordered from smallest to largest}, gives one CDF curve. The solid orange line gives the median CDF curve, while the dark (light) orange bands contain 50\% (90\%) of the CDF curves. {In more detail, for each set of three random $m_1$ draws, we record the minimum, middle and maximum value, corresponding the CDF $y$-values $1/3$, $2/3$, and $1$. Across all sets of three, we build probability distributions of the minimum, middle and maximum $m_1$ value. We summarize these probability distributions by their median (solid orange line) and central 50\% and 90\% uncertainty intervals (dark and light orange bands, respectively). }

 If the blue CDF were significantly offset from the orange, we would conclude that the BH masses in HMXBs are statistically distinct from the BBH population. However, in Fig.~\ref{fig:mass_comparison_HMXB} we see that the orange band contains the blue, such that we cannot rule out that the observed HMXB BH masses are drawn from the astrophysical BBH primary mass distribution if we correct for GW selection. 
 The consistency between the observed HMXB masses and the GW population does not imply that the two mass distributions are identical, but only that we do not have enough data to resolve the differences between the mass distributions. This is perhaps not unsurprising given the small HMXB sample size. There are only three HMXBs with confident BH masses, and this Poisson uncertainty dominates the uncertainty in the orange bands of Fig.~\ref{fig:mass_comparison_HMXB} (the orange bands also include for uncertainty in the true BBH distribution, stemming from the finite number of BBH observations, but this is subdominant to the counting uncertainty from the much smaller HMXB sample). 
 
 The consistency between BH masses in HMXBs and BBHs places limits on the importance of factors like the formation metallicity in determining the BH mass. Despite the fact that BBHs probably sample lower-metallicity formation environments than the observed HMXBs, which all formed recently at solar or super-solar metallicities, there is no observable discrepancy in the masses of the two populations once we correct for GW observational selection effects and account for the small HMXB sample size. Nevertheless, such a discrepancy may become resolvable with a larger sample of BH observations in HMXBs. Additionally, there may be HMXB observational selection effects, not accounted for here, that bias towards large BH masses and therefore cancel out the effect of metallicity.

\subsection{Black holes in low-mass X-ray binaries have similar masses to binary black holes with low-mass secondaries}
\label{sec:LMXBmass}

\begin{figure}
    \centering
    \includegraphics[width=0.5\textwidth]{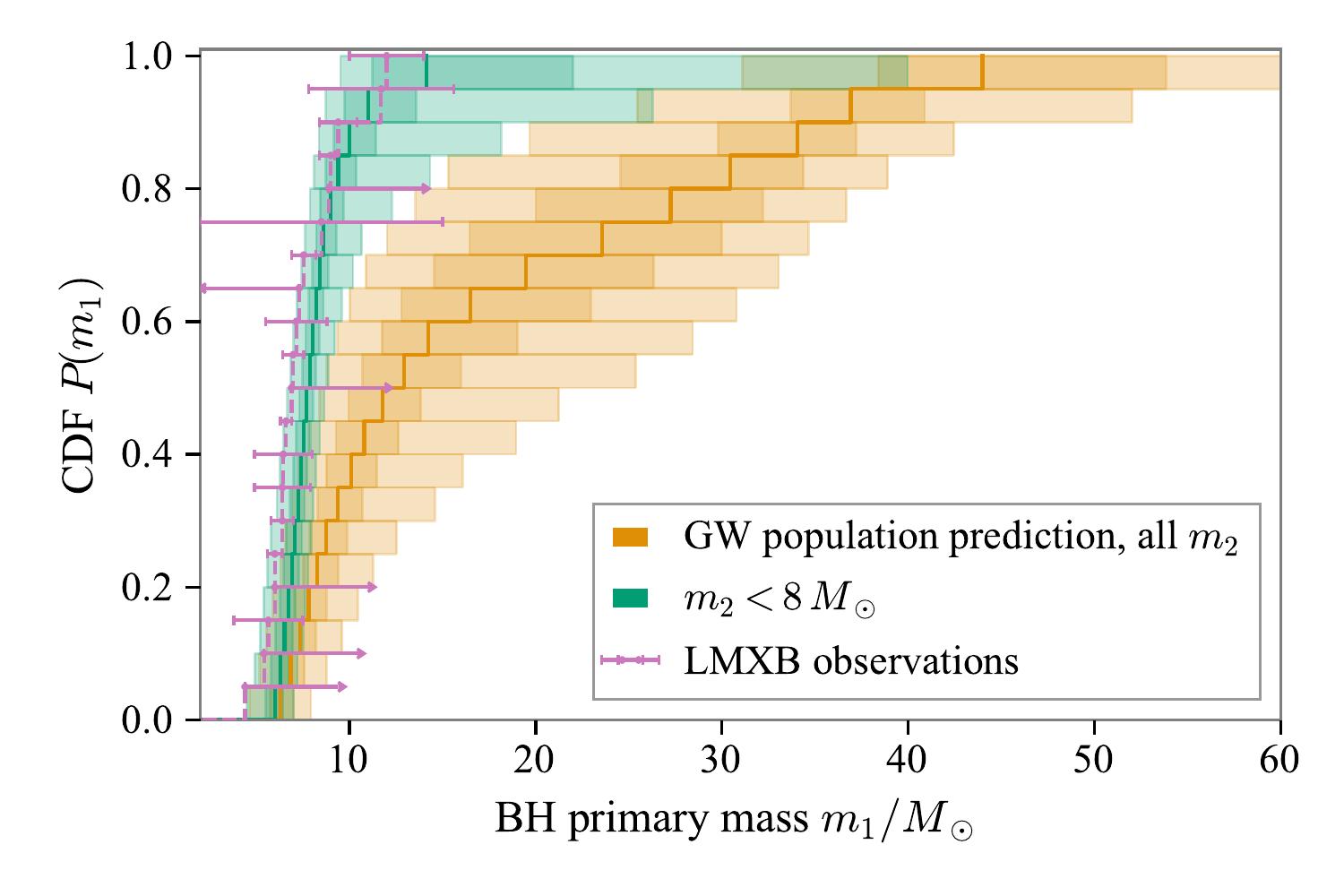}
    \caption{
    Empirical CDF for the 20 observed LMXB BH masses (pink error bars), compared to the predicted CDF of 20 random draws from the \emph{astrophysical} BBH primary mass distribution (orange bands), and the predicted CDF of 20 random draws from the BBH primary mass distribution {conditioned on $m_2 < 8\,M_\odot$} (green bands). The BHs in LMXBs are lower mass than the full BBH primary mass distribution, as seen by the offset between the pink curve and orange bands, but may be consistent with the inferred distribution of BBH primary masses with small secondary masses.}
    \label{fig:mass_comparison_LMXB_m2cuts}
\end{figure}

In Fig.~\ref{fig:mass_comparison_LMXB_m2cuts}, we carry out an analogous calculation as in Fig.~\ref{fig:mass_comparison_HMXB} of Section~\ref{sec:HMXBmass}, this time comparing the \result{20} BH masses in LMXBs (pink) to sets of \result{20} draws from the astrophysical BBH population (orange).
We draw the empirical CDF for the LMXB observations (pink dashed line) to pass through the central measured value of each observation, except in cases where only an upper or lower limit is available. In this case, we replace the error bar with an arrow, and draw the empirical CDF to pass through the available limit. From the offset between the pink curve and the orange shaded region, we see that BH masses in observed LMXBs are clearly drawn from a different population than the full BBH population. This is in agreement with the result that, without accounting for observational selection effects in either sample, BHs in LMXBs follow a different mass distribution than those in HMXBs~\citep{2011ApJ...741..103F}.  

However, we emphasize that the one-dimensional comparison between BH-XRB and BBH masses ignores the pairing between the two components in a binary~\citep[e.g.][]{2020ApJ...891L..27F}. Even if the component masses in a BBH were paired randomly, the labeling of $m_1 \geq m_2$ imposes a correlation between $m_1$ and $m_2$. Moreover, the component masses in a BBH are probably not randomly paired, in that equal-mass pairings are more common than asymmetric systems. To illustrate this, we show the inferred primary mass distribution conditioned on relatively small secondary masses, $P(m_1 \mid m_2 < 8\,M_\odot)$ as the green band in Fig.~\ref{fig:mass_comparison_LMXB_m2cuts}. We see that the green band is shifted to lower mass compared to the orange band, showing that when we consider only BBH systems with ``low" secondary masses, the primary masses also tend to be smaller. Clearly, the stellar companions to BHs in LMXBs are much lighter than any BH progenitor, and we cannot say whether the pairing between BBH components extrapolates to such low masses. However, if we believe there is some correlation between component masses and, as suggested from the BBH population, extreme mass ratios $q < 0.1$ are rare, it is not surprising that BH masses in observed LMXBs, which have very light stellar companions, are systematically smaller than BBH primary masses. In other words, the difference between LMXB and BBH masses could be driven by the mass of the secondary, rather than the mass of the BH. We see that comparing the BH masses in LMXBs to the BBHs with the smallest secondary masses $m_2 < 8\,M_\odot$ already alleviates much of the tension between the LMXB and BBH mass distribution. 

\section{Black hole spin}
\label{sec:spins}
In this section we turn to the BH spin distribution. We begin by comparing the distributions of BH spin magnitudes in HMXBs, LMXBs, and BBHs, under the assumption that in BBH systems, the component BHs are independently drawn from the same spin distribution (Section~\ref{sec:spinsiid}). We then explore an alternative model to the BBH spin distribution in which a subpopulation of systems is ``HMXB-like," with preferentially aligned, rapidly spinning primary BHs (Section~\ref{sec:spinsa1large}). The presence of such a subpopulation may explain the observed correlation between the effective inspiral spin $\chi_\mathrm{eff}$ and mass ratio $q$ pointed out by~\citet{2021arXiv210600521C}. Finally, we explore the hypothesis that LMXB spins grow by accretion. Adopting the distribution of BBH spins $a_\mathrm{GW}$ from Section~\ref{sec:spinsiid} as a proxy for the BH natal spin distribution, we infer the implied amount of spinup undergone by observed LMXBs, $a_\mathrm{LMXB} - a_\mathrm{GW}$ (Section~\ref{sec:spinup}). 

\subsection{Black holes in X-ray binaries spin faster than binary black holes}
\label{sec:spinsiid}

\begin{figure}
    \centering
    \includegraphics[width = 0.5\textwidth]{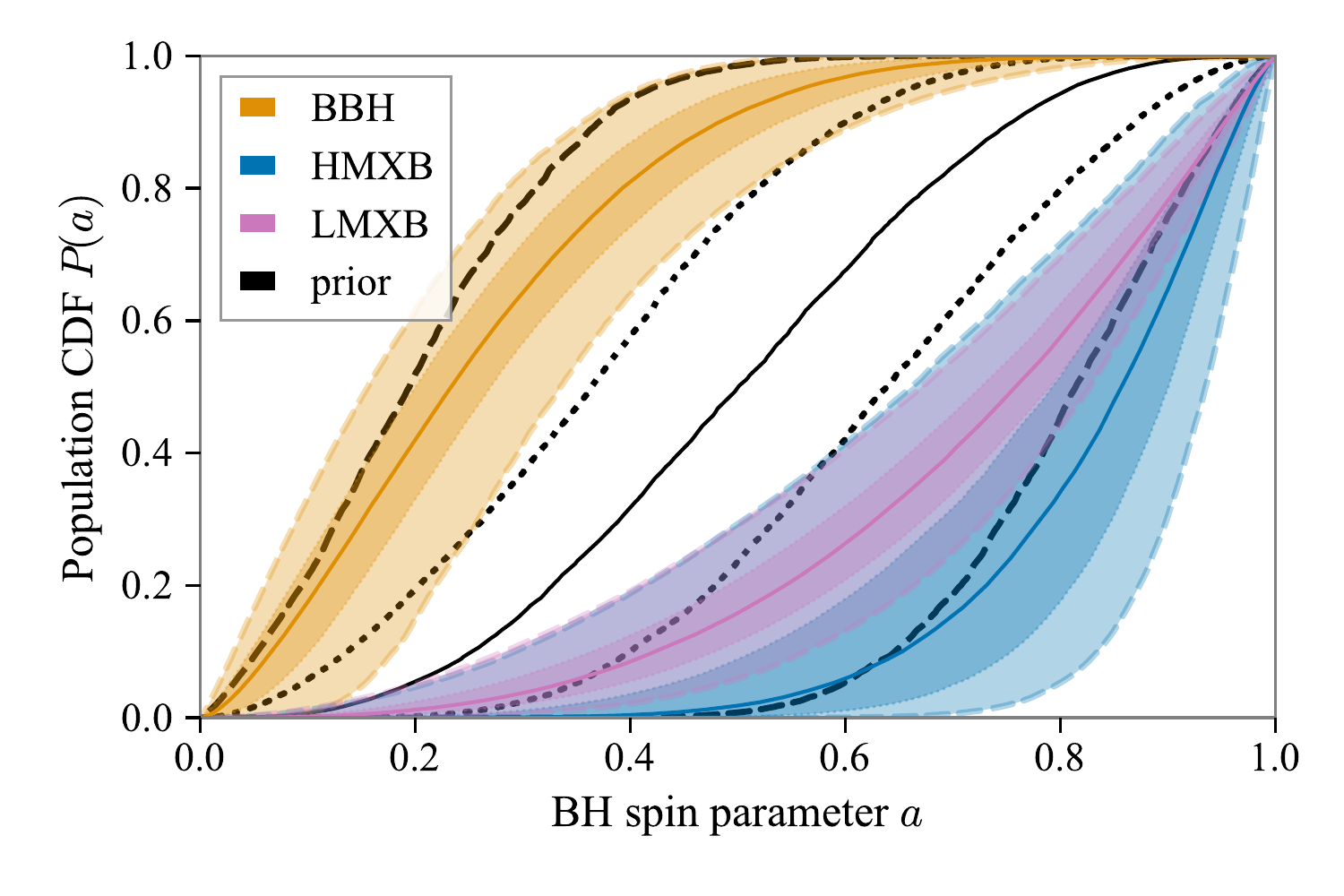}
    \caption{Inferred CDFs of the BH spin magnitude $a$ for the three populations: BBH spin distribution (orange, fit taken from~\citealt{2021ApJ...913L...7A} {in which both BH spins in a BBH are identically and independently distributed}), observed HMXB spin distribution (blue), and observed LMXB spin distribution (pink). {Following~\citealt{2021ApJ...913L...7A},} the spin magnitude distribution is fit to a beta distribution, with flat priors on the hyperparameters $\mu$ and $\sigma^2$, excluding singular beta distributions. The distribution of prior CDFs is shown by the unfilled black bands. {The BBH spin distribution is skewed to small spins, peaking at the smallest spin magnitudes allowed by the prior, whereas the observed HMXB and LMXB spin distributions have most of their support at large spins $a \gtrsim 0.4$.}}
    \label{fig:spin_pop_all}
\end{figure}

In Fig.~\ref{fig:spin_pop_all}, we compare the inferred distributions of dimensionless BH spin magnitude $a$ for BHs in BBHs, observed HMXBs, and observed LMXBs. In orange, we show the inferred BBH spin distribution from~\citet{2021ApJ...913L...7A}, which assumes that both component BHs in a binary are drawn the same spin distribution, regardless of their mass. This fit marginalizes over the spin tilt distribution, which is modeled following~\citet{2017PhRvD..96b3012T}.
In blue and pink, we show the fits to the observed HMXB and LMXB spin parameters, respectively, using the reported spins from Table~\ref{tab:BH-XRB}. 
Following the spin magnitude model that is fit to the BBHs, we assume that $p(a)$ follows a beta distribution, and restrict the two shape parameters to be greater than one to avoid singular distributions ~\citep{2019PhRvD.100d3012W,2019ApJ...882L..24A,2021ApJ...913L...7A}. 
Other than the restriction to nonsingular distributions, we take uniform priors on the mean ($\mu$) and variance ($\sigma^2$). 
Prior CDF draws are summarized by the black curves; the outer dashed curves enclose 90\% of the prior CDF draws at a given $a$, the inner dotted curves enclose 50\%, and the solid curve shows the median prior CDF at a given $a$.
We do not fit for the BH-XRB spin tilt distributions. 

We see that the inferred HMXB and LMXB spin distributions are consistent with each other within their 90\% credibility intervals, although the HMXB data allow for spin distributions that are more sharply peaked at maximal spins $a = 1$. On the other hand, the BBH spin magnitudes are significantly smaller than either the HMXB or LMXB spins. If we fit a single spin distribution to the observed LMXBs and HMXBs, the inferred posteriors on $\mu$ and $\sigma^2$ between the BH-XRBs and BBHs disagree at more than the \result{99.9\%} level.

\subsection{A subpopulation of binary black holes with highly spinning primaries?}
\label{sec:spinsa1large}

\begin{figure}
    \centering
    \includegraphics[width = 0.5\textwidth]{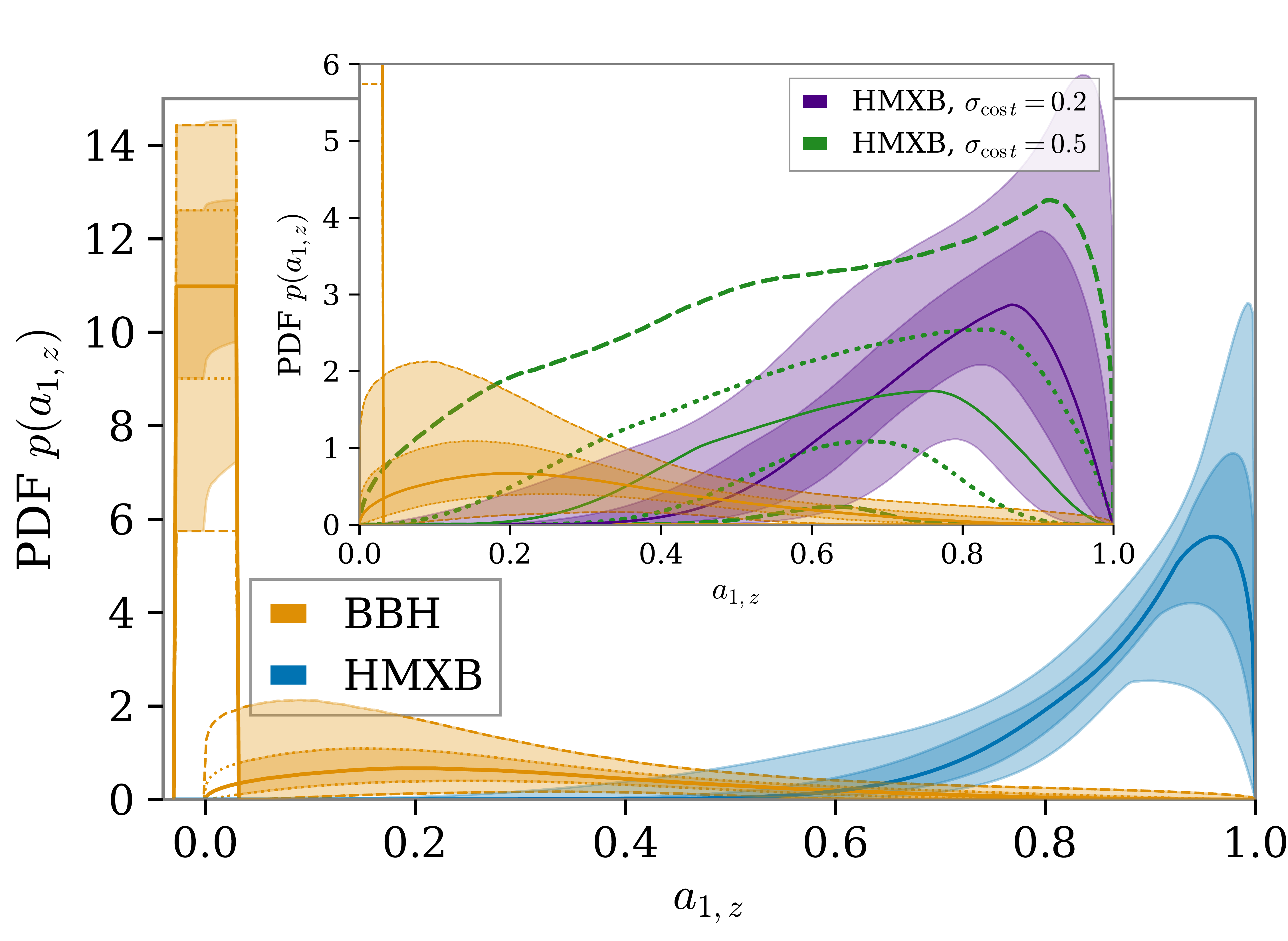}
    \caption{\emph{Main panel:} In orange, the probability distribution of the \emph{aligned} component of the \emph{primary} BH spin $a_{1,z}$ for BBH systems, inferred under the model described in Section~\ref{sec:spinsa1large} {in which the primary spin distribution consists of a nonspinning component and an aligned spin $a_{1,z} > 0$ component, while the secondary spin is fixed to zero. The two components of the primary spin distribution are shown by the unfilled dashed and dotted bands, whereas the total distribution is shown by the filled bands}. In blue, the inferred distribution of BH spins in observed HMXBs, assuming spins perfectly aligned with the orbits. \emph{Inset:} Zoom-in of the aligned $a_{1,z} > 0$ component of the BBH spin distribution (orange), compared to the observed HMXB spin distribution under different assumptions for the spin tilt distribution, with $\sigma_{\cos t} = 0.2$ (purple, filled bands) and $\sigma_{\cos t} = 0.5$ (green, unfilled).}
    \label{fig:pa1z-HMXB-inset}
\end{figure}

\begin{figure}
    \centering
    \includegraphics[width = 0.5\textwidth]{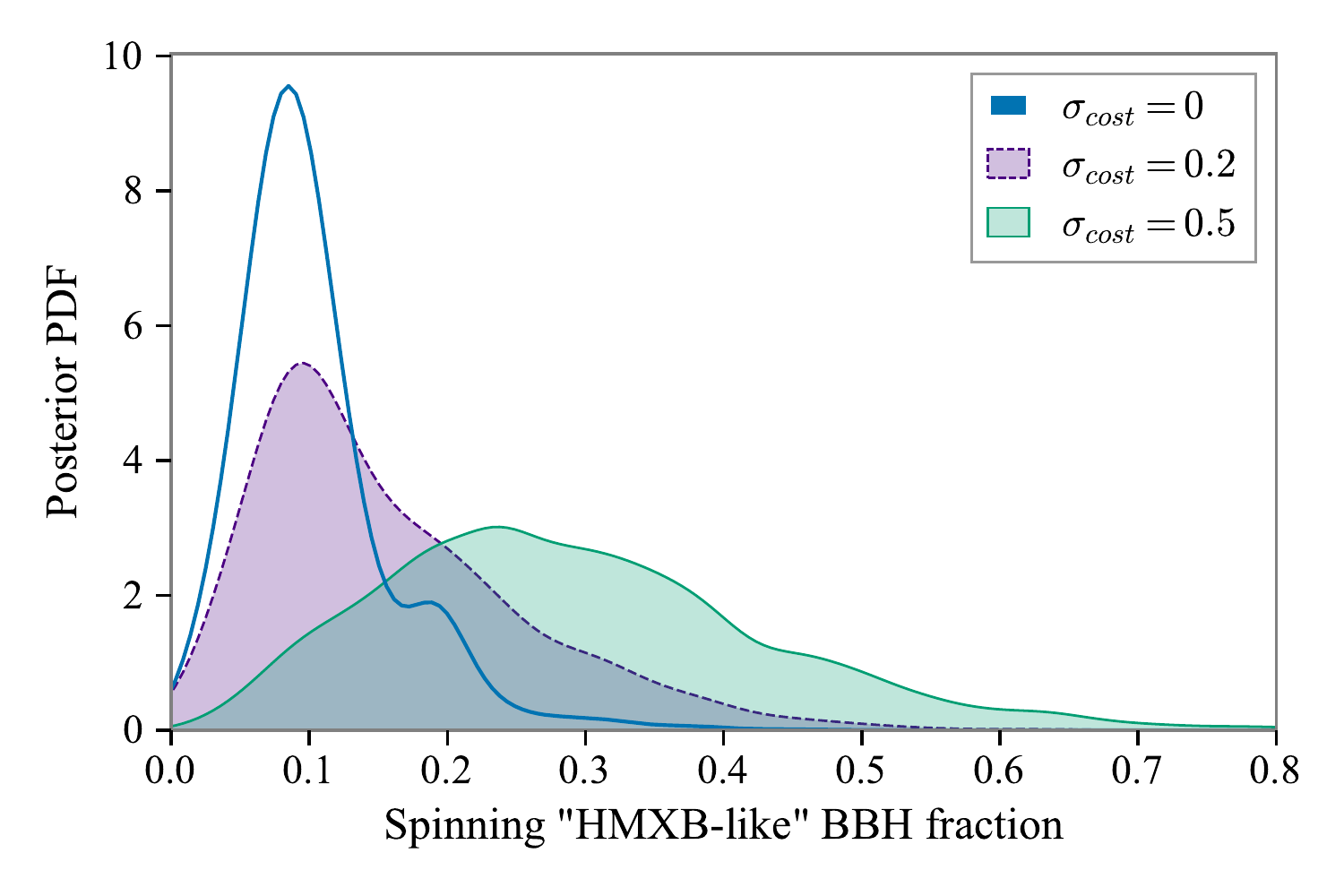}
    \caption{Fraction of BBH systems with primary spins consistent with the observed HMXB spin magnitude distribution, inferred under three different assumptions of the HMXB-like spin tilt distribution: perfect spin-orbit alignment (blue, unfilled), small possible tilts $\sigma_{\cos t} = 0.2$ (purple, dashed, filled), and possibly large tilts $\sigma_{\cos t} = 0.5$ (green, filled). The HMXB-like fractions inferred here should be considered an upper limit; as discussed in the text, our model intentionally assigns the largest possible fraction of BBH systems to have high aligned primary spins.}
    \label{fig:fraction-BBH-HMXBlike}
\end{figure}

If some fraction of BBH progenitors are similar to the observed HMXBs, we may expect a subpopulation of BBH systems with (a) nearly aligned spins, typical of isolated binary evolution~\citep{2000ApJ...541..319K,2010ApJ...719L..79F,2016ApJ...832L...2R,2017MNRAS.471.2801S,2018PhRvD..98h4036G}, and (b) at least one rapidly spinning component, which will tend to be the primary BH, because the more massive component in a BBH is usually the first-born BH. 
This is in contrast to commonly-considered evolutionary scenarios that produce a spinning second-born BH through tidal locking in the BH-He star binary phase~\citep{2016MNRAS.462..844K,2018MNRAS.473.4174Z,2018A&A...616A..28Q,2020A&A...635A..97B,2020ApJ...895L..28M,2021arXiv210906872O}. 

We therefore explore whether the BBH population 
can admit an ``HMXB-like" subpopulation of systems with nearly aligned, rapidly spinning primaries. 
Most of the spin information from GW observations comes from the well-measured, approximately conserved GW spin parameter $\chi_\mathrm{eff}$: a mass-weighted projection along the orbital angular momentum axis. Among the current BBH events, $\chi_\mathrm{eff}$ tends to be close to zero~\citep{2021ApJ...913L...7A,2021arXiv210510580R}. Small values of $\chi_\mathrm{eff}$ can most generously accommodate large, nearly-aligned spins if only one of the component BHs is spinning.
Moreover, the scenario in which some fraction of BBH systems consist of rapidly spinning primaries with nonspinning secondaries may explain the GW data, including the presence of individual events like GW190412~\citep{2020PhRvD.102d3015A,2020ApJ...899L..17Z}, GW190517~\citep{2021PhRvX..11b1053A}, and GW151226~\citep{2016PhRvL.116x1103A,2021arXiv210506486C,2021arXiv210505960M}, as well as the anti-correlation between $q$ and $\chi_\mathrm{eff}$ found in the overall BBH population~\citep{2021arXiv210600521C}.

Concretely, we fit the joint mass-redshift-spin BBH distribution:
\begin{equation}
    p(m_1, q, z, \chi_\mathrm{eff}) = p(\chi_\mathrm{eff} \mid q)p(m_1, q)p(z), 
\end{equation}
where the mass component $p(m_1, q)$ follows the \textsc{Broken Power Law} model\footnote{We omit the low-mass tapering of the mass distribution in this work, fixing $\delta_m = 0$.} and the redshift distribution follows the \textsc{Power Law Redshift} model from~\citet{2018ApJ...863L..41F,2021ApJ...913L...7A}. We model the spin distribution $p(\chi_\mathrm{eff} \mid q)$ as a mixture of two beta distribution components. One of the components is designed to contain HMXB-like systems: spin tilts smaller than 90$^{\circ}$ (implying $\chi_\mathrm{eff} > 0$) and spinning primary BHs. Because our goal is to explore the largest possible primary spins permitted by the data, we fix the secondary BHs in this component to be nonspinning, as discussed above.
We wish to absorb all the large positive $\chi_\mathrm{eff}$ systems into the first component that includes the HMXB-like systems, so we design the second $\chi_\mathrm{eff}$ component to be very tightly peaked at zero. The zero-$\chi_\mathrm{eff}$ mixture component can be thought to represent either an isotropic contribution with spin magnitudes $a = 0.03$, as we may expect for the natal spins of BHs born in isolation~\citep{2019ApJ...881L...1F} and then assembled dyanmically~\citep{2000ApJ...528L..17P}, or an approximate ``zero-spin" aligned component, implemented as a narrow peak rather than a $\delta$-function for numerical reasons as in~\citet{2021arXiv210510580R}. Our model does not include a large, negative $\chi_\mathrm{eff}$ component, because~\citet{2021arXiv210510580R} and~\citet{2021arXiv210902424G} showed that such a component is not supported by the GW data. 

Putting the ``HMXB-like" and ``zero-spin" components together, we model the spin distribution as:
\begin{align}
\label{eq:chieffmidq}
    p(\chi_\mathrm{eff} \mid q) &= (1 - f_0) \left[ \beta \left(\chi_\mathrm{eff} \bigg| \frac{\mu_{1,z}}{1+q}, \frac{\sigma_{1,z}^2}{(1+q)^2} \right)  \right] \nonumber \\ &+f_0 \left[\frac{1}{2}\beta\left(\frac{\chi_\mathrm{eff} + 1}{2} \bigg| \frac{1}{2}, \frac{\sigma^2_0(q)}{\sqrt{2}}\right)\right],
\end{align}
where $\beta(x \mid \mu, \sigma^2)$ is a beta distribution with mean $\mu$ and variance $\sigma^2$, and: 
\begin{equation}
    \sigma_0^2(q) = \frac{1 + q^2}{3(1+q)^2} a_0^2
\end{equation}
is the variance of the $\chi_\mathrm{eff}$ distribution resulting from two independently drawn isotropic spins with magnitudes $a_0$~\citep[e.g.,][]{2019MNRAS.484.4216R}, which we pick to be $a_0 = 0.03$ so that $\sigma_0^2(q) \approx 0$. 

The implied distribution of the aligned component of BBH primary spins is:
\begin{align}
\label{eq:pa1z}
    p(a_{1,z}) &= (1-f_0) \beta \left(a_{1,z} \mid \mu_{1,z}, \sigma^2_{1,z} \right) \nonumber\\ &+ f_0 \Theta (-0.03 < a_{1,z} < 0.03),
\end{align}
where $\Theta$ is an indicator function. The first term in Eq.~\ref{eq:pa1z} includes the HMXB-like systems and the second term consists of the zero-spin systems.
The orange band in the main panel of Fig.~\ref{fig:pa1z-HMXB-inset} shows the full distribution $p(a_{1,z})$ inferred under the mixture model of Eqs.~\ref{eq:chieffmidq}--\ref{eq:pa1z}, with broad, flat priors on the model hyper-parameters, excluding singular beta distributions. The inferred spin distribution of the observed BHs in HMXBs, fit to the HMXB-like component of the mixture model (i.e. fixing $f_0 = 0$), is shown in blue. 
In the main panel, we assume that BH spins in the HMXB-like component are perfectly aligned, so that the three measured HMXB spin magnitudes $a$ are identical to the aligned component $\equiv a_{1,z}$.

In the inset of Fig.~\ref{fig:pa1z-HMXB-inset}, we zoom into the HMXB-like component of the mixture model, showing the BBH fit (orange), and the HMXB fit under different assumptions for the distributions of spin tilts $t$. 
If, due to supernova kicks~\citep{2000ApJ...541..319K,2011ApJ...742...81F}, BH spins are not perfectly aligned ($a_{1,z} / a \equiv \cos t < 1$), a smaller aligned component $a_{1,z}$ can support larger spin magnitudes $a$~\citep{2017Natur.548..426F}. We follow~\citet{2017PhRvD..96b3012T} and assume that spin tilts $t$ are distributed according to a half-Gaussian in $\cos t$, peaked at $\cos t = 1$, with some standard deviation $\sigma_{\cos t}$. Allowing for small typical misalignments less than $\sigma_{\cos t} = 0.2$ ($t \sim 37^\circ$) the observed HMXB spin distribution $p(a_{1,z})$ shifts slightly to smaller values, shown by the purple band in the inset of Fig.~\ref{fig:pa1z-HMXB-inset}. If HMXB-like primary BHs have a larger spread in possible tilts, $\sigma_{\cos t} = 0.5$ ($t = 60^\circ$) we infer a broader HMXB $a_{1,z}$ distribution, shifted towards smaller aligned spins, as shown by the unfilled green band in the inset. 

In Fig.~\ref{fig:pa1z-HMXB-inset}, we showed the $p(a_{1,z})$ inference under broad priors on the HMXB-like component, which simply required positive $a_{1,z}$. Under these priors, a significant fraction of BBH systems belong to the ``zero-spin" component, $f_0 = 0.66^{+0.21}_{-0.31}$, and only $0.34^{+0.31}_{-0.21}$ of BBH systems have significant aligned primary spin, consistent with the results of~\citet{2021arXiv210902424G}. Moreover, if we insist that the HMXB-like component resembles the BH spin distribution of the observed HMXBs, the constraints on the fraction of HMXB-like systems in the BBH population tightens further. We use the inferred spin distributions of the observed HMXBs from Fig.~\ref{fig:pa1z-HMXB-inset} as a prior on the HMXB-like component, and infer the fraction $1 - f_0$ of BBH systems with HMXB-like spins. The posterior PDFs on the HMXB-like BBH fraction, under different assumptions for the spin tilt distribution of HMXB-like systems, are shown in Fig.~\ref{fig:fraction-BBH-HMXBlike}. If we define ``HMXB-like" to encompass a wide distribution of spin tilts with $\sigma_{\cos t} = 0.5$, the inferred HMXB $a_{1,z}$ distribution is less well-constrained and has support down to smaller $a_{1,z}$, as we saw in Fig.~\ref{fig:pa1z-HMXB-inset}, and so we find that up to $48\%$ of BBH systems can have HMXB-like spins (90\% credibility; green filled curve in Fig.~\ref{fig:fraction-BBH-HMXBlike}). Meanwhile, if we require a more tightly aligned spin distribution, with $\sigma_{\cos t} = 0.2$ (typical tilts less than $\sim 37^\circ$) to meet the HMXB-like criteria, the HMXB-like BBH fraction must be less than $30\%$; see the filled, purple curve in Fig.~\ref{fig:fraction-BBH-HMXBlike}. Requiring perfect alignment, we find that the HMXB-like BBH fraction is less than 19\%; see the unfilled, blue curve in Fig.~\ref{fig:fraction-BBH-HMXBlike}. These constraints should all be interpreted as upper limits on the fraction of BBH systems with spins similar to the observed HMXB spins. If secondary BHs have spin, or there is a subpopulation of BHs with intermediate spin between nonspinning and HMXB-like, we would infer a smaller HMXB-like fraction.  

{The discussion of this subsection and in particular the interpretation of Fig.~\ref{fig:pa1z-HMXB-inset} has assumed that spinning primary BHs $a_{1,z} > 0$ are similar to the BHs in HMXBs. However, the subpopulation of BBHs with strictly positive aligned primary spins $a_{1,z} > 0$ may stem from an evolutionary pathway unrelated to the observed HMXBs, such as tidal spin-up accompanied by mass inversion. In this scenario, the progenitor stellar core of the second-born BH can be spun up by tidal interactions if it is in a sufficiently tight orbit with the primary BH. This second-born BH can sometimes become the more massive primary BH in the system, yielding a subpopulation of BBHs with strictly positive aligned primary spins $a_{1,z} > 0$~\citep{2021arXiv210906872O}. If such evolutionary pathways are common, they contaminate the ``HMXB-like" subpopulation and our reported fraction of HMXB-like BBH systems is an even tighter upper limit.}

\subsection{Spinning up low-mass X-ray binaries}
\label{sec:spinup}

\begin{figure}
    \centering
    \includegraphics[width = 0.5\textwidth]{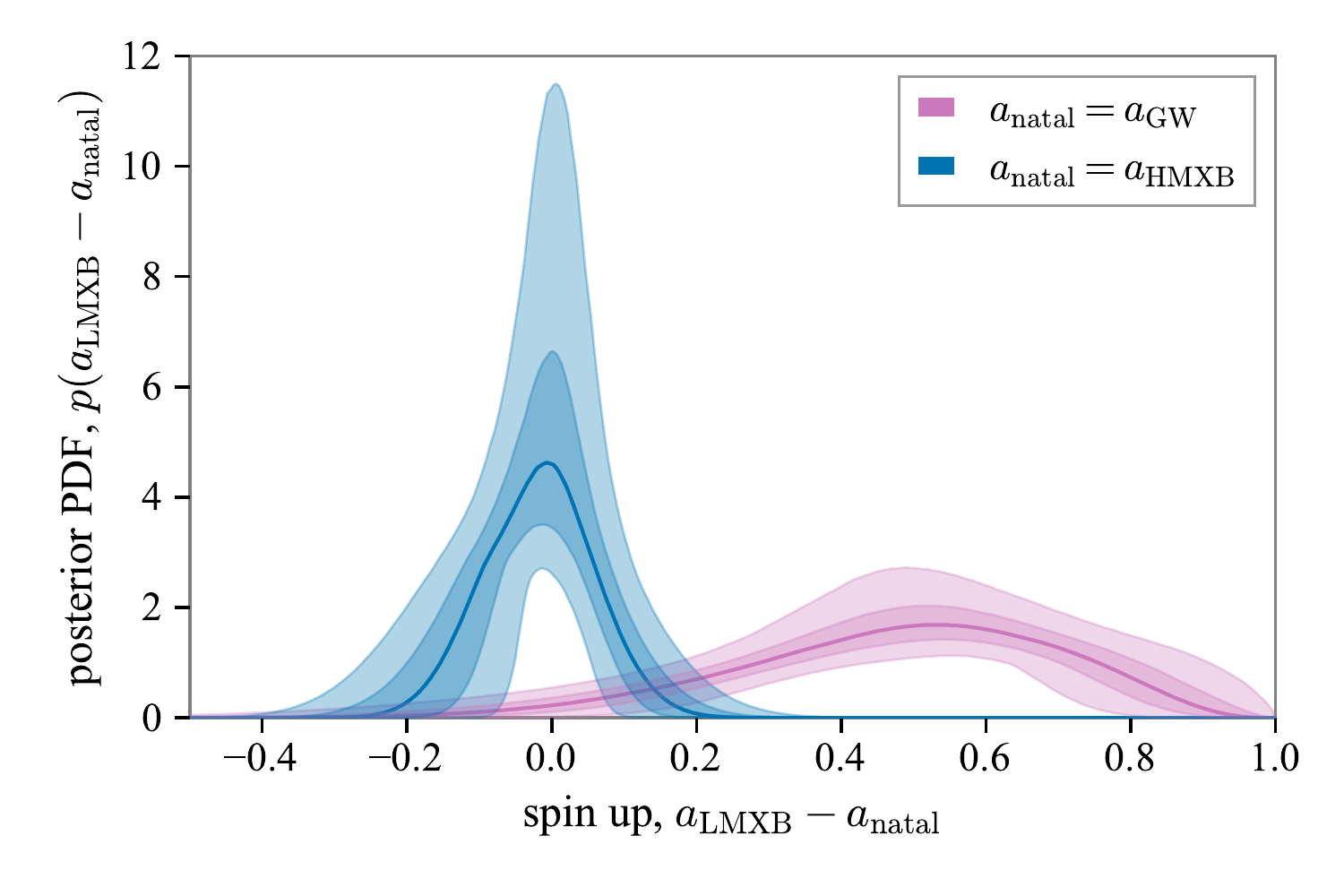}
    \caption{Posterior PDF on the amount of BH spin up in observed LMXB BHs relative to BBHs, $a_\mathrm{LMXB} - a_\mathrm{GW}$ (pink), versus the amount of LMXB spin up relative to HMXBs, $a_\mathrm{LMXB} - a_\mathrm{HMXB}$ (blue). We infer that if LMXB BHs are born with the same spin distribution as BBHs, more than \result{91\%} of observed LMXB must have been spun up with $a_\mathrm{LMXB} - a_\mathrm{GW} > 0$, gaining an average spin magnitude of \result{$0.47^{+0.10}_{-0.11}$}. On the other hand, if the observed LMXB BHs are born with the same spin distribution as the observed HMXB BHs, only a minority of systems could have experienced spin up.}
    \label{fig:spinup_LMXB}
\end{figure}

Several authors have proposed that the observed BH spins in LMXBs may not represent BH natal spins, but rather spin gained by long-term accretion from their donor star~\citep{2003MNRAS.341..385P,2005ApJ...632.1035O,2015ApJ...800...17F}. This is in contrast to BHs in wind-fed HMXBs, which cannot accrete enough material to spin up~\citep{2010Natur.468...77V,2012ApJ...747..111W}, so that we expect that the very rapid spins observed in all three HMXBs indicate their BH natal spins~\citep{2017ApJ...846L..15B,2019ApJ...870L..18Q}. 

If we assume that BBH spins, rather than HMXB spins, are representative of BH natal spins in observed LMXB systems, we can infer the amount of spin that the observed BHs in LMXBs, $a_\mathrm{LMXB}$ typically gain relative to the component spins in BBH systems, $a_\mathrm{GW}$. Using the same BBH population distribution fit by~\citet{2021ApJ...913L...7A} for $p(a_\mathrm{GW})$, we infer the distribution of the LMXB spin up, $p(a_\mathrm{LMXB} - a_\mathrm{GW})$. Assuming that the amount of LMXB spin up $a_\mathrm{LMXB} - a_\mathrm{GW}$ follows a beta distribution with uniform priors on the mean and variance, the posterior population distribution is shown in Fig.~\ref{fig:spinup_LMXB} (pink). We infer an average spinup of \result{$0.47^{+0.10}_{-0.11}$}, and find that for 90\% of distributions $p(a_\mathrm{LMXB} - a_\mathrm{GW})$ in our posterior, at least \result{91\%} of the observed LMXB BHs must have been spun up with $a_\mathrm{LMXB} - a_\mathrm{GW} > 0$. 
However, the accretion scenario in LMXBs would efficiently align the BH spin with the orbit, and thus appears to be in tension with observations of spin-orbit misalignment in microblazars like V4641 Sgr~\citep{2020MNRAS.495.2179S} and MAXI J1820+070 ~\citep{2021arXiv210907511P}. Furthermore, if the BH spins in observed HMXBs are representative of LMXB natal spins, there is no need to invoke BH spinup in observed LMXBs as the two spin distributions are consistent with one another (see Fig.~\ref{fig:spin_pop_all}). The corresponding posterior on the difference $a_\mathrm{LMXB} - a_\mathrm{HMXB}$ is shown in the blue bands of Fig.~\ref{fig:spinup_LMXB}, and is restricted to be close to zero, with an average spinup $< 0.03$. 

\section{Discussion}
X-rays and GWs have revealed distinct populations of stellar-mass BHs. A complete understanding of BH formation must explain the commonalities as well as the distinctions between BH-XRB and BBH population statistics. In the previous subsections, we compare the mass and spin distributions of the observed LMXBs and HMXBs against the astrophysical population of BBHs. Our main results are:
\begin{enumerate}
    \item The masses of BHs in observed HMXBs are consistent with the astrophysical BBH primary mass distribution, once we account for GW selection effects and the small HMXB sample size (Fig.~\ref{fig:mass_comparison_HMXB}). {In particular, with the small HMXB sample size, there is currently no evidence that differences in the formation metallicities of the two samples impart distinct BBH and HMXB mass distributions. We expect that with a larger observed sample, it would be possible to resolve differences in the mass distributions stemming from differences in formation metallicity.}
    \item The masses of BHs in observed LMXBs are significantly smaller than the astrophysical BBH primary mass distribution. However, this discrepancy may be due to a latent variable: the companion masses. When accounting for the pairing between binary components and restricting to BHs with low mass secondaries, the  observed LMXB BH masses are consistent with the BBH primary mass distribution  (Fig.~\ref{fig:mass_comparison_LMXB_m2cuts}).  
    \item The spins of BHs in observed LMXBs and HMXBs are consistent with coming from the same distribution, but are significantly faster than BBH spins (Fig.~\ref{fig:spin_pop_all}).
    \item The BBH population may include a subpopulation of systems that are similar to the observed HMXBs as far as primary spins, but, as a conservative upper limit, it must comprise no more than $\result{\sim 30\%}$ of BBH systems (Fig.~\ref{fig:fraction-BBH-HMXBlike}). 
    \item If the full BBH spin distribution represents the natal spins of the BHs in LMXBs, over \result{91\%} of observed LMXBs must have been spun up by accretion. However, if the observed BH spins in HMXBs are representative of LMXB natal spins, it is unlikely that LMXBs experience spinup (Fig.~\ref{fig:spin_pop_all}).
\end{enumerate}

We stress that our analysis does \emph{not} model observational or astrophysical selection effects in the LMXB and HMXB samples. On the other hand, our analysis accounts for observational selection effects in the BBH sample. We therefore compare the statistics of the \emph{observed} LMXB and HMXB systems against the \emph{astrophysical} BBH population as inferred from GWs. 

{One may wonder whether differences in the BH-XRB and BBH populations, specifically the BH spin distributions, stem from systematic errors in BH-XRB spin measurements. Because these measurements rely on our still-evolving understanding of XRB accretion disks, they realistically include up to a $\sim 0.1$ systematic uncertainty in addition to the statistical uncertainties quoted in Table~\ref{tab:BH-XRB}. However, as discussed in Section~\ref{sec:data}, the consistency in BH-XRB spins measured independently with reflection spectroscopy and continuum fitting for a handful of systems suggests that the systematic uncertainty for either method is not too large~\citep{doi:10.1146/annurev-astro-112420-035022}. Such modest systematic errors on BH-XRB spins would not be sufficient to bring the observed BH-XRB spin distribution into agreement with the BBH spin distribution; this would require unrealistically large systematic errors on the dimensionless spin parameter in excess of $0.5$ to reduce measured spins of $\sim 0.9$ to $\leq 0.4$.}
Instead, the distinctions between the observed LMXB and HMXB populations and the astrophysical BBH population likely result from a combination of astrophysical and observational selection. We mentioned examples of possible selection effects in Section~\ref{sec:intro}, including the different formation metallicities of the different samples, which could affect the BH masses, and the possibility of mass-dependent supernova kicks. The supernova mechanism may lead low-mass BH-XRBs, which successfully explode or experience fallback, to be (a) more easily detectable in our galaxy, but less likely to merge as BBH~\citep{2021arXiv210403596J} and (b) spinning more rapidly than their high-mass counterparts, which undergo direct collapse~\citep{2017ApJ...846L..15B,2020MNRAS.499.3214M,2020MNRAS.495.3751C,2021arXiv210407493J}. With the available data, there does not seem to be evidence for a mass-dependent BH kick distribution in LMXBs~\citep{2019MNRAS.489.3116A} or a mass-dependent spin distribution in BBHs~\citep{2021ApJ...913L...7A} as predicted by the supernova hypothesis, but if these correlations were found, they would lend support to this proposed explanation for the discrepant BH-XRB and BBH spin distributions. 

Another possible selection effect that could explain the high spins in the observed BH-XRB sample relative to the BBH spin distribution is a correlation between BH-XRB detectability and BH spin. {\citet{2021A&A...652A.138S} argued that rapidly spinning BHs in BH-XRBs are more likely to form accretion disks that lead to longer and brighter X-ray emission.} Furthermore, a correlation between BH spins and X-ray detectability may arise if scenarios for spinning up the BH or its progenitor stellar core are more efficient at small orbital periods and BH-XRBs in tight orbits are more detectable~\citep{2019ApJ...870L..18Q}. For example,~\citet{2021arXiv210803774H} argued that HMXB systems are observable only if the companion main sequence star is close to filling its Roche lobe, which may in turn imply that the system is less likely to survive a common envelope and become a merging BBH system~\citep{2021ApJ...908..118N}.

 Further detailed modeling of the evolutionary histories of HMXBs, LMXBs and BBHs will shed insight into the astrophysical and observational effects that distinguish these populations~\citep{Liotine,Siegel}.

\acknowledgments
We gratefully acknowledge helpful discussions with Christopher Berry, Christopher Reynolds, and with the organizers and participants of the EAS 2021 Session ``The Birth, Life and Death of Black Holes." We thank Reed Essick and Jose Ezquiaga for their comments on the manuscript. M.F. is supported by NASA through NASA Hubble Fellowship grant HST-HF2-51455.001-A awarded by the Space Telescope Science Institute, which is operated by the Association of Universities for
Research in Astronomy, Incorporated, under NASA contract NAS5-26555.
M.F. is grateful for the hospitality of Perimeter Institute where part of this work was carried out.
Research at Perimeter Institute is supported in part by the Government of Canada through the
Department of Innovation, Science and Economic Development Canada and by the Province of
Ontario through the Ministry of Economic Development, Job Creation and Trade.
V.K. is grateful for support from a Guggenheim Fellowship, from CIFAR as a Senior Fellow, and from Northwestern University, including the Daniel I. Linzer Distinguished University Professorship fund. 

\bibliographystyle{aasjournal}
\bibliography{references}

\begin{thebibliography}{}
\expandafter\ifx\csname natexlab\endcsname\relax\def\natexlab#1{#1}\fi
\providecommand{\url}[1]{\href{#1}{#1}}
\providecommand{\dodoi}[1]{doi:~\href{http://doi.org/#1}{\nolinkurl{#1}}}
\providecommand{\doeprint}[1]{\href{http://ascl.net/#1}{\nolinkurl{http://ascl.net/#1}}}
\providecommand{\doarXiv}[1]{\href{https://arxiv.org/abs/#1}{\nolinkurl{https://arxiv.org/abs/#1}}}

\bibitem[{{Aasi} {et~al.}(2015){Aasi}, {Abbott}, {Abbott}, {Abbott},
  {Abernathy}, {Ackley}, {Adams}, {Adams}, {Addesso}, \&
  et~al.}]{2015CQGra..32g4001L}
{Aasi}, J., {Abbott}, B.~P., {Abbott}, R., {et~al.} 2015, Classical and Quantum
  Gravity, 32, 074001, \dodoi{10.1088/0264-9381/32/7/074001}

\bibitem[{{Abbott} {et~al.}(2016{\natexlab{a}}){Abbott}, {Abbott}, {Abbott},
  {Abernathy}, {Acernese}, {Ackley}, {Adams}, {Adams}, {Addesso}, {Adhikari},
  \& et~al.}]{2016PhRvL.116f1102A}
{Abbott}, B.~P., {Abbott}, R., {Abbott}, T.~D., {et~al.} 2016{\natexlab{a}},
  \prl, 116, 061102, \dodoi{10.1103/PhysRevLett.116.061102}

\bibitem[{{Abbott} {et~al.}(2016{\natexlab{b}}){Abbott}, {Abbott}, {Abbott},
  {Abernathy}, {Acernese}, {Ackley}, {Adams}, {Adams}, {Addesso}, {Adhikari},
  \& et~al.}]{2016PhRvX...6d1015A}
---. 2016{\natexlab{b}}, Physical Review X, 6, 041015,
  \dodoi{10.1103/PhysRevX.6.041015}

\bibitem[{{Abbott} {et~al.}(2016{\natexlab{c}}){Abbott}, {Abbott}, {Abbott},
  {Abernathy}, {Acernese}, {Ackley}, {Adams}, {Adams}, {Addesso}, {Adhikari},
  \& et~al.}]{2016PhRvL.116x1103A}
---. 2016{\natexlab{c}}, \prl, 116, 241103,
  \dodoi{10.1103/PhysRevLett.116.241103}

\bibitem[{{Abbott} {et~al.}(2019{\natexlab{a}}){Abbott}, {Abbott}, {Abbott},
  {Abraham}, {Acernese}, {Ackley}, {Adams}, {Adhikari}, {Adya}, {Affeldt}, \&
  et~al.}]{2019PhRvX...9c1040A}
---. 2019{\natexlab{a}}, Physical Review X, 9, 031040,
  \dodoi{10.1103/PhysRevX.9.031040}

\bibitem[{{Abbott} {et~al.}(2019{\natexlab{b}}){Abbott}, {Abbott}, {Abbott},
  {Abraham}, {Acernese}, {Ackley}, {Adams}, {Adhikari}, {Adya}, {Affeldt}, \&
  et~al.}]{2019ApJ...882L..24A}
---. 2019{\natexlab{b}}, \apjl, 882, L24, \dodoi{10.3847/2041-8213/ab3800}

\bibitem[{{Abbott} {et~al.}(2020){Abbott}, {Abbott}, {Abraham}, {Acernese},
  {Ackley}, {Adams}, {Adhikari}, {Adya}, {Affeldt}, {Agathos}, \&
  et~al.}]{2020PhRvD.102d3015A}
{Abbott}, R., {Abbott}, T.~D., {Abraham}, S., {et~al.} 2020, \prd, 102, 043015,
  \dodoi{10.1103/PhysRevD.102.043015}

\bibitem[{{Abbott} {et~al.}(2021{\natexlab{a}}){Abbott}, {Abbott}, {Abraham},
  {Acernese}, {Ackley}, {Adams}, {Adams}, {Adhikari}, {Adya}, {Affeldt}, \&
  et~al.}]{2021PhRvX..11b1053A}
---. 2021{\natexlab{a}}, Physical Review X, 11, 021053,
  \dodoi{10.1103/PhysRevX.11.021053}

\bibitem[{{Abbott} {et~al.}(2021{\natexlab{b}}){Abbott}, {Abbott}, {Acernese},
  {Ackley}, {Adams}, {Adhikari}, {Adhikari}, {Adya}, \&
  et~al.}]{2021arXiv210801045T}
{Abbott}, R., {Abbott}, T.~D., {Acernese}, F., {et~al.} 2021{\natexlab{b}},
  arXiv e-prints, arXiv:2108.01045.
\newblock \doarXiv{2108.01045}

\bibitem[{{Abbott} {et~al.}(2021{\natexlab{c}}){Abbott}, {Abbott}, {Abraham},
  {Acernese}, {Ackley}, {Adams}, {Adams}, {Adhikari}, {Adya}, {Affeldt}, \&
  et~al.}]{2021ApJ...913L...7A}
{Abbott}, R., {Abbott}, T.~D., {Abraham}, S., {et~al.} 2021{\natexlab{c}},
  \apjl, 913, L7, \dodoi{10.3847/2041-8213/abe949}

\bibitem[{{Acernese} {et~al.}(2015){Acernese}, {Agathos}, {Agatsuma}, {Aisa},
  {Allemandou}, {Allocca}, {Amarni}, {Astone}, {Balestri}, {Ballardin}, \&
  et~al.}]{2015CQGra..32b4001A}
{Acernese}, F., {Agathos}, M., {Agatsuma}, K., {et~al.} 2015, Classical and
  Quantum Gravity, 32, 024001, \dodoi{10.1088/0264-9381/32/2/024001}

\bibitem[{{Atri} {et~al.}(2019){Atri}, {Miller-Jones}, {Bahramian}, {Plotkin},
  {Jonker}, {Nelemans}, {Maccarone}, {Sivakoff}, {Deller}, {Chaty}, \&
  et~al.}]{2019MNRAS.489.3116A}
{Atri}, P., {Miller-Jones}, J.~C.~A., {Bahramian}, A., {et~al.} 2019, \mnras,
  489, 3116, \dodoi{10.1093/mnras/stz2335}

\bibitem[{{Batta} {et~al.}(2017){Batta}, {Ramirez-Ruiz}, \&
  {Fryer}}]{2017ApJ...846L..15B}
{Batta}, A., {Ramirez-Ruiz}, E., \& {Fryer}, C. 2017, \apjl, 846, L15,
  \dodoi{10.3847/2041-8213/aa8506}

\bibitem[{{Bavera} {et~al.}(2020){Bavera}, {Fragos}, {Qin}, {Zapartas},
  {Neijssel}, {Mandel}, {Batta}, {Gaebel}, {Kimball}, \&
  {Stevenson}}]{2020A&A...635A..97B}
{Bavera}, S.~S., {Fragos}, T., {Qin}, Y., {et~al.} 2020, \aap, 635, A97,
  \dodoi{10.1051/0004-6361/201936204}

\bibitem[{{Belczynski} {et~al.}(2011){Belczynski}, {Bulik}, \&
  {Bailyn}}]{2011ApJ...742L...2B}
{Belczynski}, K., {Bulik}, T., \& {Bailyn}, C. 2011, \apjl, 742, L2,
  \dodoi{10.1088/2041-8205/742/1/L2}

\bibitem[{{Belczynski} {et~al.}(2012){Belczynski}, {Bulik}, \&
  {Fryer}}]{2012arXiv1208.2422B}
{Belczynski}, K., {Bulik}, T., \& {Fryer}, C.~L. 2012, arXiv e-prints,
  arXiv:1208.2422.
\newblock \doarXiv{1208.2422}

\bibitem[{{Callister} {et~al.}(2021){Callister}, {Haster}, {Ng}, {Vitale}, \&
  {Farr}}]{2021arXiv210600521C}
{Callister}, T.~A., {Haster}, C.-J., {Ng}, K. K.~Y., {Vitale}, S., \& {Farr},
  W.~M. 2021, arXiv e-prints, arXiv:2106.00521.
\newblock \doarXiv{2106.00521}

\bibitem[{{Casares} \& {Jonker}(2014)}]{2014SSRv..183..223C}
{Casares}, J., \& {Jonker}, P.~G. 2014, \ssr, 183, 223,
  \dodoi{10.1007/s11214-013-0030-6}

\bibitem[{{Chan} {et~al.}(2020){Chan}, {M{\"u}ller}, \&
  {Heger}}]{2020MNRAS.495.3751C}
{Chan}, C., {M{\"u}ller}, B., \& {Heger}, A. 2020, \mnras, 495, 3751,
  \dodoi{10.1093/mnras/staa1431}

\bibitem[{{Chia} {et~al.}(2021){Chia}, {Olsen}, {Roulet}, {Dai}, {Venumadhav},
  {Zackay}, \& {Zaldarriaga}}]{2021arXiv210506486C}
{Chia}, H.~S., {Olsen}, S., {Roulet}, J., {et~al.} 2021, arXiv e-prints,
  arXiv:2105.06486.
\newblock \doarXiv{2105.06486}

\bibitem[{{Corral-Santana} {et~al.}(2016){Corral-Santana}, {Casares},
  {Mu{\~n}oz-Darias}, {Bauer}, {Mart{\'\i}nez-Pais}, \&
  {Russell}}]{2016A&A...587A..61C}
{Corral-Santana}, J.~M., {Casares}, J., {Mu{\~n}oz-Darias}, T., {et~al.} 2016,
  \aap, 587, A61, \dodoi{10.1051/0004-6361/201527130}

\bibitem[{{Corral-Santana} {et~al.}(2015){Corral-Santana}, {Casares},
  {Munoz-Darias}, {Bauer}, {Martinez-Pais}, \& {Russell}}]{2015yCat..35870061C}
{Corral-Santana}, J.~M., {Casares}, J., {Munoz-Darias}, T., {et~al.} 2015,
  VizieR Online Data Catalog, J/A+A/587/A61

\bibitem[{{Falanga} {et~al.}(2021){Falanga}, {Bakala}, {La Placa}, {De Falco},
  {De Rosa}, \& {Stella}}]{2021MNRAS.504.3424F}
{Falanga}, M., {Bakala}, P., {La Placa}, R., {et~al.} 2021, \mnras, 504, 3424,
  \dodoi{10.1093/mnras/stab1147}

\bibitem[{{Farr} {et~al.}(2011{\natexlab{a}}){Farr}, {Kremer}, {Lyutikov}, \&
  {Kalogera}}]{2011ApJ...742...81F}
{Farr}, W.~M., {Kremer}, K., {Lyutikov}, M., \& {Kalogera}, V.
  2011{\natexlab{a}}, \apj, 742, 81, \dodoi{10.1088/0004-637X/742/2/81}

\bibitem[{{Farr} {et~al.}(2011{\natexlab{b}}){Farr}, {Sravan}, {Cantrell},
  {Kreidberg}, {Bailyn}, {Mandel}, \& {Kalogera}}]{2011ApJ...741..103F}
{Farr}, W.~M., {Sravan}, N., {Cantrell}, A., {et~al.} 2011{\natexlab{b}}, \apj,
  741, 103, \dodoi{10.1088/0004-637X/741/2/103}

\bibitem[{{Farr} {et~al.}(2017){Farr}, {Stevenson}, {Miller}, {Mandel}, {Farr},
  \& {Vecchio}}]{2017Natur.548..426F}
{Farr}, W.~M., {Stevenson}, S., {Miller}, M.~C., {et~al.} 2017, \nat, 548, 426,
  \dodoi{10.1038/nature23453}

\bibitem[{{Fishbach} \& {Holz}(2020)}]{2020ApJ...891L..27F}
{Fishbach}, M., \& {Holz}, D.~E. 2020, \apjl, 891, L27,
  \dodoi{10.3847/2041-8213/ab7247}

\bibitem[{{Fishbach} {et~al.}(2018){Fishbach}, {Holz}, \&
  {Farr}}]{2018ApJ...863L..41F}
{Fishbach}, M., {Holz}, D.~E., \& {Farr}, W.~M. 2018, \apjl, 863, L41,
  \dodoi{10.3847/2041-8213/aad800}

\bibitem[{{Fragos} \& {McClintock}(2015)}]{2015ApJ...800...17F}
{Fragos}, T., \& {McClintock}, J.~E. 2015, \apj, 800, 17,
  \dodoi{10.1088/0004-637X/800/1/17}

\bibitem[{{Fragos} {et~al.}(2010){Fragos}, {Tremmel}, {Rantsiou}, \&
  {Belczynski}}]{2010ApJ...719L..79F}
{Fragos}, T., {Tremmel}, M., {Rantsiou}, E., \& {Belczynski}, K. 2010, \apjl,
  719, L79, \dodoi{10.1088/2041-8205/719/1/L79}

\bibitem[{{Fuller} \& {Ma}(2019)}]{2019ApJ...881L...1F}
{Fuller}, J., \& {Ma}, L. 2019, \apjl, 881, L1,
  \dodoi{10.3847/2041-8213/ab339b}

\bibitem[{{Galaudage} {et~al.}(2021){Galaudage}, {Talbot}, {Nagar}, {Jain},
  {Thrane}, \& {Mandel}}]{2021arXiv210902424G}
{Galaudage}, S., {Talbot}, C., {Nagar}, T., {et~al.} 2021, arXiv e-prints,
  arXiv:2109.02424.
\newblock \doarXiv{2109.02424}

\bibitem[{{Gerosa} {et~al.}(2018){Gerosa}, {Berti}, {O'Shaughnessy},
  {Belczynski}, {Kesden}, {Wysocki}, \& {Gladysz}}]{2018PhRvD..98h4036G}
{Gerosa}, D., {Berti}, E., {O'Shaughnessy}, R., {et~al.} 2018, \prd, 98,
  084036, \dodoi{10.1103/PhysRevD.98.084036}

\bibitem[{{Giesers} {et~al.}(2018){Giesers}, {Dreizler}, {Husser}, {Kamann},
  {Anglada Escud{\'e}}, {Brinchmann}, {Carollo}, {Roth}, {Weilbacher}, \&
  {Wisotzki}}]{2018MNRAS.475L..15G}
{Giesers}, B., {Dreizler}, S., {Husser}, T.-O., {et~al.} 2018, \mnras, 475,
  L15, \dodoi{10.1093/mnrasl/slx203}

\bibitem[{{Hirai} \& {Mandel}(2021)}]{2021arXiv210803774H}
{Hirai}, R., \& {Mandel}, I. 2021, arXiv e-prints, arXiv:2108.03774.
\newblock \doarXiv{2108.03774}

\bibitem[{{Janka} {et~al.}(2021){Janka}, {Wongwathanarat}, \&
  {Kramer}}]{2021arXiv210407493J}
{Janka}, H.~T., {Wongwathanarat}, A., \& {Kramer}, M. 2021, arXiv e-prints,
  arXiv:2104.07493.
\newblock \doarXiv{2104.07493}

\bibitem[{{Jonker} {et~al.}(2021){Jonker}, {Kaur}, {Stone}, \&
  {Torres}}]{2021arXiv210403596J}
{Jonker}, P.~G., {Kaur}, K., {Stone}, N., \& {Torres}, M. A.~P. 2021, arXiv
  e-prints, arXiv:2104.03596.
\newblock \doarXiv{2104.03596}

\bibitem[{{Kalogera}(2000)}]{2000ApJ...541..319K}
{Kalogera}, V. 2000, \apj, 541, 319, \dodoi{10.1086/309400}

\bibitem[{{Kushnir} {et~al.}(2016){Kushnir}, {Zaldarriaga}, {Kollmeier}, \&
  {Waldman}}]{2016MNRAS.462..844K}
{Kushnir}, D., {Zaldarriaga}, M., {Kollmeier}, J.~A., \& {Waldman}, R. 2016,
  \mnras, 462, 844, \dodoi{10.1093/mnras/stw1684}

\bibitem[{{Laycock} {et~al.}(2015){Laycock}, {Maccarone}, \&
  {Christodoulou}}]{2015MNRAS.452L..31L}
{Laycock}, S. G.~T., {Maccarone}, T.~J., \& {Christodoulou}, D.~M. 2015,
  \mnras, 452, L31, \dodoi{10.1093/mnrasl/slv082}

\bibitem[{{LIGO Scientific Collaboration} \& {Virgo
  Collaboration}(2021)}]{ligo_scientific_collaboration_and_virgo_2021_748570}
{LIGO Scientific Collaboration}, \& {Virgo Collaboration}. 2021, Test GWTC-2
  Data Set,  Zenodo, \dodoi{10.5072/zenodo.748570}

\bibitem[{{Liotine} {et~al.}(2021)}]{Liotine}
{Liotine}, C., {et~al.} 2021, \emph{in prep.}

\bibitem[{{Loredo}(2004)}]{2004AIPC..735..195L}
{Loredo}, T.~J. 2004, in American Institute of Physics Conference Series, Vol.
  735, Bayesian Inference and Maximum Entropy Methods in Science and
  Engineering: 24th International Workshop on Bayesian Inference and Maximum
  Entropy Methods in Science and Engineering, ed. R.~{Fischer}, R.~{Preuss}, \&
  U.~V. {Toussaint}, 195--206, \dodoi{10.1063/1.1835214}

\bibitem[{{Lu} {et~al.}(2016){Lu}, {Sinukoff}, {Ofek}, {Udalski}, \&
  {Kozlowski}}]{2016ApJ...830...41L}
{Lu}, J.~R., {Sinukoff}, E., {Ofek}, E.~O., {Udalski}, A., \& {Kozlowski}, S.
  2016, \apj, 830, 41, \dodoi{10.3847/0004-637X/830/1/41}

\bibitem[{{Mandel} {et~al.}(2019){Mandel}, {Farr}, \&
  {Gair}}]{2019MNRAS.486.1086M}
{Mandel}, I., {Farr}, W.~M., \& {Gair}, J.~R. 2019, \mnras, 486, 1086,
  \dodoi{10.1093/mnras/stz896}

\bibitem[{{Mandel} \& {Fragos}(2020)}]{2020ApJ...895L..28M}
{Mandel}, I., \& {Fragos}, T. 2020, \apjl, 895, L28,
  \dodoi{10.3847/2041-8213/ab8e41}

\bibitem[{{Mandel} \& {M{\"u}ller}(2020)}]{2020MNRAS.499.3214M}
{Mandel}, I., \& {M{\"u}ller}, B. 2020, \mnras, 499, 3214,
  \dodoi{10.1093/mnras/staa3043}

\bibitem[{{Mateu-Lucena} {et~al.}(2021){Mateu-Lucena}, {Husa}, {Colleoni},
  {Estell{\'e}s}, {Garc{\'\i}a-Quir{\'o}s}, {Keitel}, {de Lluc Planas}, \&
  {Ramos-Buades}}]{2021arXiv210505960M}
{Mateu-Lucena}, M., {Husa}, S., {Colleoni}, M., {et~al.} 2021, arXiv e-prints,
  arXiv:2105.05960.
\newblock \doarXiv{2105.05960}

\bibitem[{{Miller} \& {Miller}(2015)}]{2015PhR...548....1M}
{Miller}, M.~C., \& {Miller}, J.~M. 2015, \physrep, 548, 1,
  \dodoi{10.1016/j.physrep.2014.09.003}

\bibitem[{{Miller-Jones} {et~al.}(2021){Miller-Jones}, {Bahramian}, {Orosz},
  {Mandel}, {Gou}, {Maccarone}, {Neijssel}, {Zhao}, {Zi{\'o}{\l}kowski},
  {Reid}, {Uttley}, {Zheng}, {Byun}, {Dodson}, {Grinberg}, {Jung}, {Kim},
  {Marcote}, {Markoff}, {Rioja}, {Rushton}, {Russell}, {Sivakoff}, {Tetarenko},
  {Tudose}, \& {Wilms}}]{2021Sci...371.1046M}
{Miller-Jones}, J. C.~A., {Bahramian}, A., {Orosz}, J.~A., {et~al.} 2021,
  Science, 371, 1046, \dodoi{10.1126/science.abb3363}

\bibitem[{{Motta} {et~al.}(2021){Motta}, {Rodriguez}, {Jourdain}, {Del Santo},
  {Belanger}, {Cangemi}, {Grinberg}, {Kajava}, {Kuulkers}, {Malzac}, \&
  et~al.}]{2021NewAR..9301618M}
{Motta}, S.~E., {Rodriguez}, J., {Jourdain}, E., {et~al.} 2021, \nar, 93,
  101618, \dodoi{10.1016/j.newar.2021.101618}

\bibitem[{{Neijssel} {et~al.}(2021){Neijssel}, {Vinciguerra},
  {Vigna-G{\'o}mez}, {Hirai}, {Miller-Jones}, {Bahramian}, {Maccarone}, \&
  {Mandel}}]{2021ApJ...908..118N}
{Neijssel}, C.~J., {Vinciguerra}, S., {Vigna-G{\'o}mez}, A., {et~al.} 2021,
  \apj, 908, 118, \dodoi{10.3847/1538-4357/abde4a}

\bibitem[{{Nitz} {et~al.}(2019){Nitz}, {Capano}, {Nielsen}, {Reyes}, {White},
  {Brown}, \& {Krishnan}}]{2019ApJ...872..195N}
{Nitz}, A.~H., {Capano}, C., {Nielsen}, A.~B., {et~al.} 2019, \apj, 872, 195,
  \dodoi{10.3847/1538-4357/ab0108}

\bibitem[{{Nitz} {et~al.}(2021){Nitz}, {Capano}, {Kumar}, {Wang}, {Kastha},
  {Sch{\"a}fer}, {Dhurkunde}, \& {Cabero}}]{2021arXiv210509151N}
{Nitz}, A.~H., {Capano}, C.~D., {Kumar}, S., {et~al.} 2021, arXiv e-prints,
  arXiv:2105.09151.
\newblock \doarXiv{2105.09151}

\bibitem[{{Nitz} {et~al.}(2020){Nitz}, {Dent}, {Davies}, {Kumar}, {Capano},
  {Harry}, {Mozzon}, {Nuttall}, {Lundgren}, \&
  {T{\'a}pai}}]{2020ApJ...891..123N}
{Nitz}, A.~H., {Dent}, T., {Davies}, G.~S., {et~al.} 2020, \apj, 891, 123,
  \dodoi{10.3847/1538-4357/ab733f}

\bibitem[{{Olejak} \& {Belczynski}(2021)}]{2021arXiv210906872O}
{Olejak}, A., \& {Belczynski}, K. 2021, arXiv e-prints, arXiv:2109.06872.
\newblock \doarXiv{2109.06872}

\bibitem[{{O'Shaughnessy} {et~al.}(2005){O'Shaughnessy}, {Kaplan}, {Kalogera},
  \& {Belczynski}}]{2005ApJ...632.1035O}
{O'Shaughnessy}, R., {Kaplan}, J., {Kalogera}, V., \& {Belczynski}, K. 2005,
  \apj, 632, 1035, \dodoi{10.1086/444346}

\bibitem[{{Perna} {et~al.}(2019){Perna}, {Wang}, {Farr}, {Leigh}, \&
  {Cantiello}}]{2019ApJ...878L...1P}
{Perna}, R., {Wang}, Y.-H., {Farr}, W.~M., {Leigh}, N., \& {Cantiello}, M.
  2019, \apjl, 878, L1, \dodoi{10.3847/2041-8213/ab2336}

\bibitem[{{Podsiadlowski} {et~al.}(2003){Podsiadlowski}, {Rappaport}, \&
  {Han}}]{2003MNRAS.341..385P}
{Podsiadlowski}, P., {Rappaport}, S., \& {Han}, Z. 2003, \mnras, 341, 385,
  \dodoi{10.1046/j.1365-8711.2003.06464.x}

\bibitem[{{Portegies Zwart} \& {McMillan}(2000)}]{2000ApJ...528L..17P}
{Portegies Zwart}, S.~F., \& {McMillan}, S. L.~W. 2000, \apjl, 528, L17,
  \dodoi{10.1086/312422}

\bibitem[{{Poutanen} {et~al.}(2021){Poutanen}, {Veledina}, {Berdyugin},
  {Berdyugina}, {Jermak}, {Jonker}, {Kajava}, {Kosenkov}, {Kravtsov},
  {Piirola}, \& et~al.}]{2021arXiv210907511P}
{Poutanen}, J., {Veledina}, A., {Berdyugin}, A.~V., {et~al.} 2021, arXiv
  e-prints, arXiv:2109.07511.
\newblock \doarXiv{2109.07511}

\bibitem[{{Qin} {et~al.}(2018){Qin}, {Fragos}, {Meynet}, {Andrews},
  {S{\o}rensen}, \& {Song}}]{2018A&A...616A..28Q}
{Qin}, Y., {Fragos}, T., {Meynet}, G., {et~al.} 2018, \aap, 616, A28,
  \dodoi{10.1051/0004-6361/201832839}

\bibitem[{{Qin} {et~al.}(2019){Qin}, {Marchant}, {Fragos}, {Meynet}, \&
  {Kalogera}}]{2019ApJ...870L..18Q}
{Qin}, Y., {Marchant}, P., {Fragos}, T., {Meynet}, G., \& {Kalogera}, V. 2019,
  \apjl, 870, L18, \dodoi{10.3847/2041-8213/aaf97b}

\bibitem[{{Reis} {et~al.}(2013){Reis}, {Reynolds}, {Miller}, {Walton},
  {Maitra}, {King}, \& {Degenaar}}]{2013ApJ...778..155R}
{Reis}, R.~C., {Reynolds}, M.~T., {Miller}, J.~M., {et~al.} 2013, \apj, 778,
  155, \dodoi{10.1088/0004-637X/778/2/155}

\bibitem[{{Remillard} \& {McClintock}(2006)}]{2006ARA&A..44...49R}
{Remillard}, R.~A., \& {McClintock}, J.~E. 2006, \araa, 44, 49,
  \dodoi{10.1146/annurev.astro.44.051905.092532}

\bibitem[{Reynolds(2021)}]{doi:10.1146/annurev-astro-112420-035022}
Reynolds, C.~S. 2021, Annual Review of Astronomy and Astrophysics, 59, 117,
  \dodoi{10.1146/annurev-astro-112420-035022}

\bibitem[{{Rodriguez} {et~al.}(2016){Rodriguez}, {Zevin}, {Pankow}, {Kalogera},
  \& {Rasio}}]{2016ApJ...832L...2R}
{Rodriguez}, C.~L., {Zevin}, M., {Pankow}, C., {Kalogera}, V., \& {Rasio},
  F.~A. 2016, \apjl, 832, L2, \dodoi{10.3847/2041-8205/832/1/L2}

\bibitem[{{Roulet} {et~al.}(2021){Roulet}, {Chia}, {Olsen}, {Dai},
  {Venumadhav}, {Zackay}, \& {Zaldarriaga}}]{2021arXiv210510580R}
{Roulet}, J., {Chia}, H.~S., {Olsen}, S., {et~al.} 2021, arXiv e-prints,
  arXiv:2105.10580.
\newblock \doarXiv{2105.10580}

\bibitem[{{Roulet} {et~al.}(2020){Roulet}, {Venumadhav}, {Zackay}, {Dai}, \&
  {Zaldarriaga}}]{2020PhRvD.102l3022R}
{Roulet}, J., {Venumadhav}, T., {Zackay}, B., {Dai}, L., \& {Zaldarriaga}, M.
  2020, \prd, 102, 123022, \dodoi{10.1103/PhysRevD.102.123022}

\bibitem[{{Roulet} \& {Zaldarriaga}(2019)}]{2019MNRAS.484.4216R}
{Roulet}, J., \& {Zaldarriaga}, M. 2019, \mnras, 484, 4216,
  \dodoi{10.1093/mnras/stz226}

\bibitem[{{Salvesen} \& {Miller}(2021)}]{2021MNRAS.500.3640S}
{Salvesen}, G., \& {Miller}, J.~M. 2021, \mnras, 500, 3640,
  \dodoi{10.1093/mnras/staa3325}

\bibitem[{{Salvesen} \& {Pokawanvit}(2020)}]{2020MNRAS.495.2179S}
{Salvesen}, G., \& {Pokawanvit}, S. 2020, \mnras, 495, 2179,
  \dodoi{10.1093/mnras/staa1094}

\bibitem[{{Schneider} {et~al.}(2021){Schneider}, {Podsiadlowski}, \&
  {M{\"u}ller}}]{2021A&A...645A...5S}
{Schneider}, F.~R.~N., {Podsiadlowski}, P., \& {M{\"u}ller}, B. 2021, \aap,
  645, A5, \dodoi{10.1051/0004-6361/202039219}

\bibitem[{{Sen} {et~al.}(2021){Sen}, {Xu}, {Langer}, {El Mellah},
  {Sch{\"u}rmann}, \& {Quast}}]{2021A&A...652A.138S}
{Sen}, K., {Xu}, X.~T., {Langer}, N., {et~al.} 2021, \aap, 652, A138,
  \dodoi{10.1051/0004-6361/202141214}

\bibitem[{{Siegel} {et~al.}(2021)}]{Siegel}
{Siegel}, J., {et~al.} 2021, \emph{in prep.}

\bibitem[{{Stevenson} {et~al.}(2017){Stevenson}, {Berry}, \&
  {Mandel}}]{2017MNRAS.471.2801S}
{Stevenson}, S., {Berry}, C. P.~L., \& {Mandel}, I. 2017, \mnras, 471, 2801,
  \dodoi{10.1093/mnras/stx1764}

\bibitem[{{Talbot} \& {Thrane}(2017)}]{2017PhRvD..96b3012T}
{Talbot}, C., \& {Thrane}, E. 2017, \prd, 96, 023012,
  \dodoi{10.1103/PhysRevD.96.023012}

\bibitem[{{Talbot} \& {Thrane}(2018)}]{2018ApJ...856..173T}
---. 2018, \apj, 856, 173, \dodoi{10.3847/1538-4357/aab34c}

\bibitem[{{Tauris} \& {van den Heuvel}(2006)}]{2006csxs.book..623T}
{Tauris}, T.~M., \& {van den Heuvel}, E.~P.~J. 2006, {Formation and evolution
  of compact stellar X-ray sources}, Vol.~39, 623--665

\bibitem[{{Taylor} \& {Reynolds}(2018)}]{2018ApJ...855..120T}
{Taylor}, C., \& {Reynolds}, C.~S. 2018, \apj, 855, 120,
  \dodoi{10.3847/1538-4357/aaad63}

\bibitem[{{Thompson} {et~al.}(2019){Thompson}, {Kochanek}, {Stanek}, {Badenes},
  {Post}, {Jayasinghe}, {Latham}, {Bieryla}, {Esquerdo}, {Berlind}, \&
  et~al.}]{2019Sci...366..637T}
{Thompson}, T.~A., {Kochanek}, C.~S., {Stanek}, K.~Z., {et~al.} 2019, Science,
  366, 637, \dodoi{10.1126/science.aau4005}

\bibitem[{{Thrane} \& {Talbot}(2019)}]{2019PASA...36...10T}
{Thrane}, E., \& {Talbot}, C. 2019, \pasa, 36, e010,
  \dodoi{10.1017/pasa.2019.2}

\bibitem[{{Valsecchi} {et~al.}(2010){Valsecchi}, {Glebbeek}, {Farr}, {Fragos},
  {Willems}, {Orosz}, {Liu}, \& {Kalogera}}]{2010Natur.468...77V}
{Valsecchi}, F., {Glebbeek}, E., {Farr}, W.~M., {et~al.} 2010, \nat, 468, 77,
  \dodoi{10.1038/nature09463}

\bibitem[{{Venumadhav} {et~al.}(2020){Venumadhav}, {Zackay}, {Roulet}, {Dai},
  \& {Zaldarriaga}}]{2020PhRvD.101h3030V}
{Venumadhav}, T., {Zackay}, B., {Roulet}, J., {Dai}, L., \& {Zaldarriaga}, M.
  2020, \prd, 101, 083030, \dodoi{10.1103/PhysRevD.101.083030}

\bibitem[{{Vink} {et~al.}(2021){Vink}, {Higgins}, {Sander}, \&
  {Sabhahit}}]{2021MNRAS.504..146V}
{Vink}, J.~S., {Higgins}, E.~R., {Sander}, A. A.~C., \& {Sabhahit}, G.~N. 2021,
  \mnras, 504, 146, \dodoi{10.1093/mnras/stab842}

\bibitem[{{Vitale} {et~al.}(2020){Vitale}, {Gerosa}, {Farr}, \&
  {Taylor}}]{2020arXiv200705579V}
{Vitale}, S., {Gerosa}, D., {Farr}, W.~M., \& {Taylor}, S.~R. 2020, arXiv
  e-prints, arXiv:2007.05579.
\newblock \doarXiv{2007.05579}

\bibitem[{{Wong} {et~al.}(2012){Wong}, {Valsecchi}, {Fragos}, \&
  {Kalogera}}]{2012ApJ...747..111W}
{Wong}, T.-W., {Valsecchi}, F., {Fragos}, T., \& {Kalogera}, V. 2012, \apj,
  747, 111, \dodoi{10.1088/0004-637X/747/2/111}

\bibitem[{{Wyrzykowski} \& {Mandel}(2020)}]{2020A&A...636A..20W}
{Wyrzykowski}, {\L}., \& {Mandel}, I. 2020, \aap, 636, A20,
  \dodoi{10.1051/0004-6361/201935842}

\bibitem[{{Wysocki} {et~al.}(2019){Wysocki}, {Lange}, \&
  {O'Shaughnessy}}]{2019PhRvD.100d3012W}
{Wysocki}, D., {Lange}, J., \& {O'Shaughnessy}, R. 2019, \prd, 100, 043012,
  \dodoi{10.1103/PhysRevD.100.043012}

\bibitem[{{Zackay} {et~al.}(2019){Zackay}, {Venumadhav}, {Dai}, {Roulet}, \&
  {Zaldarriaga}}]{2019PhRvD.100b3007Z}
{Zackay}, B., {Venumadhav}, T., {Dai}, L., {Roulet}, J., \& {Zaldarriaga}, M.
  2019, \prd, 100, 023007, \dodoi{10.1103/PhysRevD.100.023007}

\bibitem[{{Zaldarriaga} {et~al.}(2018){Zaldarriaga}, {Kushnir}, \&
  {Kollmeier}}]{2018MNRAS.473.4174Z}
{Zaldarriaga}, M., {Kushnir}, D., \& {Kollmeier}, J.~A. 2018, \mnras, 473,
  4174, \dodoi{10.1093/mnras/stx2577}

\bibitem[{{Zevin} {et~al.}(2020){Zevin}, {Berry}, {Coughlin}, {Chatziioannou},
  \& {Vitale}}]{2020ApJ...899L..17Z}
{Zevin}, M., {Berry}, C. P.~L., {Coughlin}, S., {Chatziioannou}, K., \&
  {Vitale}, S. 2020, \apjl, 899, L17, \dodoi{10.3847/2041-8213/aba8ef}

\end{thebibliography}
\end{document}